\documentclass[sigconf, nonacm]{acmart}

\usepackage{subfigure}
\usepackage{xcolor}
\usepackage{xspace}
\usepackage{booktabs}
\usepackage{url}
\usepackage{multirow}
\usepackage{listings}
\usepackage{makecell}
\usepackage{tabularx}
\usepackage{utfsym}
\usepackage[linesnumbered, boxed]{algorithm2e}
\usepackage{amsmath, amsthm}
\usepackage{pifont}

\usepackage{graphicx}
\usepackage{mylstlinebgrd}

\newcommand\revision[1]{\textcolor{black}{#1}\xspace}
\newcommand\delete[1]{\textcolor{black}{#1}\xspace}

\newcommand\figref[1]{Fig.~\ref{#1}}

\newcommand\lstref[1]{Listing~\ref{#1}}
\newcommand\tabref[1]{Table~\ref{#1}}

\newcommand\secref[1]{Sec.~\ref{#1}}

\newcommand\appref[1]{Appendix~\ref{#1}}
\newcommand\algref[1]{Alg.~\ref{#1}}

\newcommand{\nlsql}{Text-to-SQL\xspace}

\newcommand{\bsprompt}{Basic Prompt\xspace}
\newcommand{\textprompt}{Text Representation Prompt\xspace}
\newcommand{\sqlprompt}{Code Representation Prompt\xspace}
\newcommand{\openaiprompt}{OpenAI Demostration Prompt\xspace}
\newcommand{\alpacaprompt}{Alpaca SFT Prompt\xspace}

\newcommand{\abbsprompt}{BS\,$_P$\xspace}
\newcommand{\abtextprompt}{TR\,$_P$\xspace}
\newcommand{\absqlprompt}{CR\,$_P$\xspace}
\newcommand{\abopenaiprompt}{OD\,$_P$\xspace}
\newcommand{\abalpacaprompt}{AS\,$_P$\xspace}

\newcommand{\qsselector}{Question Similarity Selection\xspace}
\newcommand{\slmselector}{Masked Question Similarity Selection\xspace}
\newcommand{\qrsselector}{Query Similarity Selection\xspace}
\newcommand{\pqsselector}{DAIL Selection\xspace}

\newcommand{\abrandselector}{Random\xspace}
\newcommand{\abqsselector}{QTS\,$_S$\xspace}
\newcommand{\abslmselector}{MQS\,$_S$\xspace}
\newcommand{\abqrsselector}{QRS\,$_S$\xspace}
\newcommand{\abpqsselector}{DAIL\,$_S$\xspace}

\newcommand{\absqlorg}{SO\,$_O$\xspace}
\newcommand{\abpairorg}{DAIL\,$_O$\xspace}
\newcommand{\abfiorg}{FI\,$_O$\xspace}

\newcommand{\sqlorg}{SQL-Only Organization\xspace}
\newcommand{\pairorg}{DAIL Organization\xspace}
\newcommand{\fiorg}{Full-Information Organization\xspace}
\newcommand{\ours}{DAIL-SQL\xspace}

\newcommand{\gptfour}{GPT-4\xspace}
\newcommand{\chatgpt}{GPT-3.5-TURBO\xspace}
\newcommand{\davinci}{TEXT-DAVINCI-003\xspace}
\newcommand{\vicuna}{Vicuna-33B\xspace}

\newcommand{\wrule}{With Rule\xspace}
\newcommand{\worule}{Without Rule\xspace}
\newcommand{\wfk}{With Foreign Keys\xspace}

\definecolor{eclipseBlue}{RGB}{42,0.0,255}
\definecolor{eclipseGreen}{RGB}{63,180,95}
\definecolor{eclipsePurple}{RGB}{175,0,25}
\definecolor{codewhite}{rgb}{0.70,0.70,0.70}
\definecolor{codegray}{rgb}{0.5,0.5,0.5}
\definecolor{codepurple}{rgb}{0.58,0,0.82}
\definecolor{backcolour}{rgb}{0.95,0.95,0.92}

\definecolor{instructioncolor}{rgb}{0.6275, 0.7686, 0.6157}
\definecolor{taskcolor}{rgb}{0.8824, 0.9255, 0.7843}
\definecolor{prefixcolor}{rgb}{0.9922, 1, 0.6824}

\lstdefinelanguage{Prompt}{
	backgroundcolor=\color{backcolour},   
	keywordstyle=\color{magenta},
	numberstyle=\tiny\color{codegray},
	basicstyle=\ttfamily\footnotesize,
	breakatwhitespace=false,         
	breaklines=true,   
    breakindent=-3.5pt,
	captionpos=b,                    
	keepspaces=true,                 
	numbers=left,                    
	numbersep=5pt,                  
	showspaces=false,                
	showstringspaces=false,
	showtabs=false,                  
	tabsize=4,
	escapeinside={`}{`},
	morecomment = [s][\color{eclipseGreen}\bfseries]{How}{?},
        morecomment = [l][\color{eclipseBlue}\bfseries]{SELECT},
        morecomment = [l][\color{eclipsePurple}\bfseries]{\$\{DATABASE_SCHEMA\}},
        morecomment = [s][\color{eclipsePurple}\bfseries]{CREATE}{;},
        morecomment = [l][\color{eclipsePurple}\bfseries]{Table},
        morecomment = [l][\color{eclipsePurple}\bfseries]{continents},
        morecomment = [l][\color{eclipsePurple}\bfseries]{countries},
        morecomment = [l][\color{codewhite}\bfseries]{\$\{TARGET_QUESTION\}},
    postbreak={
       \mbox{
           \lst@linebreakbgrd
           \rotatebox[y=0.7ex]{180}{\color{black}$\Lsh\,$}
       }
    },
}

\lstdefinelanguage{Question}{
	backgroundcolor=\color{backcolour},   
	sensitive = true,
	morecomment = [s]{Gold}{SQL},
	commentstyle ={\color{red}\bfseries\underbar},
	morestring = [b]",
	morestring = [b]',
	stringstyle = \color{eclipseGreen},
	basicstyle=\ttfamily\footnotesize,
	breaklines=true,
	alsoletter=!?-,
	emph={StableVicuna-13B, Vicuna-13B, Gold-SQL, Write a sql to answer},
	emphstyle={\color{red}\bfseries\underbar},
	captionpos=b,
	escapeinside={[}{]},
	keepspaces=true,              
	showspaces=false,                
	showstringspaces=false,
	showtabs=false,                  
	tabsize=4,
	columns=flexible
}

\begin{document}

\title{\nlsql Empowered by Large Language Models: A Benchmark Evaluation}

\author{Dawei Gao$^*$}
\affiliation{%
  \institution{Alibaba Group}
}
\email{gaodawei.gdw@alibaba-inc.com}

\author{Haibin Wang$^*$}
\affiliation{%
  \institution{Alibaba Group}
}
\email{binke.whb@alibaba-inc.com}

\author{Yaliang Li}
\affiliation{%
  \institution{Alibaba Group}
}
\email{yaliang.li@alibaba-inc.com}

\author{Xiuyu Sun}
\affiliation{%
  \institution{Alibaba Group}
}
\email{xiuyu.sxy@alibaba-inc.com}

\author{Yichen Qian}
\affiliation{%
  \institution{Alibaba Group}
}
\email{yichen.qyc@alibaba-inc.com}

\author{Bolin Ding}
\affiliation{%
  \institution{Alibaba Group}
}
\email{bolin.ding@alibaba-inc.com}

\author{Jingren Zhou}
\affiliation{%
  \institution{Alibaba Group}
}
\email{jingren.zhou@alibaba-inc.com}

\begin{abstract}
Large language models (LLMs) have emerged as a new paradigm for \nlsql task. 
However, the absence of a systematical benchmark inhibits the development of designing effective, efficient and economic LLM-based \nlsql solutions. 
To address this challenge, in this paper, we first conduct a systematical and extensive comparison over existing prompt engineering methods, including question representation, example selection and example organization, and with these experimental results, we elaborate their pros and cons. 
Based on these findings, we propose a new integrated solution, named \ours, which refreshes the Spider leaderboard with $86.6\%$ execution accuracy and sets a new bar. 

To explore the potential of open-source LLM, we investigate them in various scenarios, and further enhance their performance with supervised fine-tuning. 
Our explorations highlight open-source LLMs' potential in \nlsql, as well as the advantages and disadvantages of the supervised fine-tuning. 
Additionally, towards an efficient and economic LLM-based \nlsql solution, we emphasize the token efficiency in prompt engineering and compare the prior studies under this metric. 
We hope that our work provides a deeper understanding of \nlsql with LLMs, and inspires further investigations and broad applications.
\end{abstract}
\maketitle
\renewcommand*{\thefootnote}{\fnsymbol{footnote}}
\footnotetext[1]{Co-first authors.}

\section{Introduction}
\label{sec:intro}

\nlsql, as one challenging task in both natural language processing and database communities, maps natural language questions on the given relational database into SQL queries~\cite{nl2sql1, nl2sql2}. 
Most previous works~\cite{rat-sql, li2023graphix, li2023resdsql, DBLP:conf/acl/HuiGWQLLSL22, DBLP:conf/acl/ZhengWDWL22} focus on extracting the question-to-SQL patterns and generalizing them by training an encoder-decoder model with \nlsql corpus. 
In recent years, large language models (LLMs) have emerged as a new paradigm for \nlsql~\cite{DBLP:journals/corr/abs-2303-13547, rajkumar2022evaluating, trummer2022codexdb}. 
Notably, equipped with GPT-4~\cite{gpt4}, Pourreza et al.~\cite{din-sql} achieved the first place in Spider leaderboard~\cite{leaderboard} with $85.3\%$ execution accuracy.
Different from prior studies, the core problem in LLM-based \nlsql solution is how to prompt LLM to generate correct SQL queries, namely prompt engineering. 
Such prompt engineering involves question representations~\cite{openaiprompt,dong2023c3,DBLP:journals/corr/abs-2305-11853,din-sql}, examples selection~\cite{DBLP:conf/acl-deelio/LiuSZDCC22, DBLP:journals/corr/abs-2304-13301, enhancing}, and example organization~\cite{DBLP:journals/corr/abs-2304-13301}.

\textbf{\nlsql prompt engineering needs a systematic study.} 
Although prior studies have made remarkable progress, there still lacks a systematic study for prompt engineering in LLM-based \nlsql solutions. 
Specifically, for question representation, most existing research textualize structured knowledge as schema, and further add task instructions and foreign keys to form prompts~\cite{DBLP:journals/corr/abs-2305-01598, enhancing}. 
Besides, some studies~\cite{enhancing, DBLP:journals/corr/abs-2305-11853} represent tables as several ``\textit{CREATE TABLE}'' SQL statements, and prompt LLMs to answer the target question in comments. 
However, even with similar representation, their detailed task instructions can lead to significant performance gap. 
For example, in OpenAI's official \nlsql demo~\cite{openaiprompt}, they employ the pound sign ``$\#$'' to differentiate prompt from response, yielding an impressive performance~\cite{DBLP:journals/corr/abs-2303-13547}; If such a sign is removed, the performance will significantly drop.
Therefore, there are burgeoning demands for a systematic study over different representations and examine how to work well with LLMs. 
Regarding example selection, a common practice is to encode the most similar examples in the same representation with the target question~\cite{enhancing, DBLP:journals/corr/abs-2305-11853, DBLP:journals/corr/abs-2303-13547}. 
Nan et al.~\cite{enhancing} further underline the importance of diversity in example selection. 
While for organization, most prior studies represent examples with full information, including instruction, schema, question and ground truth SQL queries. 
Besides, Guo et al.~\cite{DBLP:journals/corr/abs-2304-13301} only keep SQL queries in the selected examples to guide the LLM with less tokens. 
Together with different LLMs' preferences, the optimal selection and organization strategies in LLM-based \nlsql solution remain ambiguous.  
Therefore, a systematical study on prompt engineering, spanning different LLMs, question representations, example selection and organizations, is highly anticipated.

\textbf{The potential of open-source LLMs is underexplored.} 
% Why exploring open-source LLMs is very important? 
Very recently, open-source LLMs are constantly expanding and show remarkable advancement in programming, mathematical reasoning, and text generation tasks. 
% What's the current status? 
However, previous \nlsql research primarily focuses on OpenAI LLMs, leaving open-source LLMs unstudied. 
Besides, compared with OpenAI LLMs, open-source ones generally have limited functionality in understanding context and generating coherent response. 
% What's the conclusion?
Thus, a critical challenge for open-source LLMs is to further enhance their performance in \nlsql, which can be achieved by supervised fine-tuning.

\textbf{Prompt efficiency remains a challenging open question.} 
% turn to efficiency
In LLM-based \nlsql, another critical challenge is efficiency. 
% why it's important
The reason is that most prior studies focus on OpenAI LLMs, and calling their APIs are expensive, time-consuming and restricted in rate limits~\cite{ratelimites}, especially for in-context learning prompts with multiple examples. 
% how about the current status? 
However, the prior studies may not well tackle this challenge. 
Specifically, based on the observed inverted-U shape in execution accuracy with respect to prompt length, Chang et al.~\cite{DBLP:journals/corr/abs-2305-11853} conjectures that LLMs may have a sweet spot in terms of prompt length, but leaves exploring efficient prompt engineering a challenging open question.

In light of above challenges, we focus on providing a comprehensive, systematical and fair benchmark for LLM-based \nlsql. 
Specifically, our benchmark discusses both the effectiveness and efficiency of various prompt engineering strategies, as well as the feasibility of open-source LLMs.
They are detailed as follows.

To provide a systematical and in-depth understanding of \nlsql prompt engineering, we empirically evaluate several strategies from prior studies. 
First, we compare several typical question representations in zero-shot scenario with different LLMs, and identify their pros and cons.
After that, we investigate example selection and organization strategies in few-shot scenario. 
For example selection, we compare different selection strategies and further verify the hypothesis that LLMs learn from the mappings between question and SQL skeleton. 
Regarding example organization, we explore the option of displaying full information, solely SQL queries or question-SQL pair. 

After that, we highlight the potential of open-source LLMs in both in-context learning and supervised fine-tuning. 
Specifically, we empirically study various open-source LLMs with different prompt engineering strategies, and observe the significant benefits of increasing scale of LLMs and having a good alignment \cite{instructgpt}. 
To further enhance their performance, we fine-tune and evaluate open-source LLMs using various representations. 
With this comparison, we demonstrate that similar to in-context learning, representation strategy is also critical for supervised fine-tuning. 
These explorations underline the potential of an effective solution for \nlsql.
Moreover, after fine-tuning we also observe a decrease in in-context learning capability, which requires further study. 
We believe these explorations will benefit practical \nlsql applications.

Towards a more economic and efficient solution, we further evaluate different strategies in terms of token efficiency. 
Such evaluation aims at searching for a cost-effective strategy, which is supposed to achieve considerable performance with less tokens. 
To fulfill such goal, we consider token efficiency in the whole process of prompt engineering, including choices for question representation, example selection and organization.

Last but not least, our integrated solution, named \ours, refreshes the Spider leaderboard with $86.6\%$ execution accuracy, and wins the first place. 
Compared with previous solutions, \ours encodes structure knowledge as SQL statements, selects examples based on their skeleton similarities and removes cross-domain knowledge from examples for token efficiency. 
Before \ours, the state-of-the-art performance in the Spider leaderboard is $85.3\%$~\cite{din-sql}. 
Therefore, our solution sets a new bar, and hope our comprehensive study will inspire more further works.

\textbf{Contribution} Our main contributions and results are summarized as follows: 
\begin{itemize} 
    \item \revision{We systematically study prompt engineering for LLM-based \nlsql methods, including five question representations, two prompt components, four example selections, and three example organizations on four LLMs. The study sheds light on identifying suitable question representations and key points to leverage the in-context learning capacity of LLMs for \nlsql task.}
    \item \revision{To the best of our knowledge, we are the first to explore open-source LLMs for both in-context learning and supervised fine-tuning for \nlsql task. We provide insights into the potential of the open-source LLMs by employing SFT for \nlsql task.}
    \item \revision{We also empirically compare different prompts in terms of cost efficiency, which provides practical guidance for real-world \nlsql applications.} 	
    \item \revision{Last but not least, we propose a new solution, named \ours, which successes in leveraging the in-context learning capacity of LLMs and achieving a balance between performance and token efficiency. Notably, it refreshes the Spider leaderboard with $86.6\%$ execution accuracy, which surpasses the best state-of-the-art solution by $1.3\%$ with much less token cost.}
\end{itemize}

\section{Preliminary}
\label{sec:preliminary}

\nlsql aims at automatically translating natural language questions into SQL queries. 
It bridges the gap between non-expert users and database systems, greatly improves the efficiency of data processing, and contributes to a wider range of applications such as intelligent database service, automatic data analysis and database question-answering.  
However, \nlsql is still a quiet challenging task, due to the difficulty in fully understanding natural language questions and generating correct SQL queries~\cite{DBLP:journals/corr/abs-2208-13629,nl2sql1}.

Extensive studies of \nlsql have been conducted in both database and natural language processing communities. 
\revision{
Some early studies tackle \nlsql task with pre-defined rules or query enumeration~\cite{DBLP:journals/pvldb/SenLQOEDSMSS20, DBLP:conf/sigmod/BaikJCJ20, quamar2022natural}, or treat it as a sequence-to-sequence task, focusing on training machine learning models with an encoder-decoder architecture~\cite{DBLP:conf/ijcai/CaiXZYLL18, DBLP:conf/coling/PopescuMVYKS22, qi2022rasat}.
With rapid advancement of deep learning, numerous techniques are applied to help \nlsql task, such as attention mechanism~\cite{DBLP:conf/ijcnn/LiuSZWLK23}, graph representation~\cite{DBLP:conf/emnlp/XuWWFS18, li2023graphix, DBLP:conf/acl/ZhengWDWL22, rat-sql, qi2022rasat, DBLP:conf/acl/HuiGWQLLSL22}, syntax parsing~\cite{DBLP:conf/acl/GuoZGXLLZ19, scholak2021picard, li2023resdsql, wang2022proton}, etc. 
One of the most representative is BERT~\cite{bert}, which has been widely used in \nlsql and achieved SOTA performances at that time~\cite{valuenet, tabert}. 
}
Besides, to narrow the gap between \nlsql research and its real-world deployment, numerous large-scale benchmark datasets have been released, including WikiSQL~\cite{wikisql}, Spider~\cite{spider}, KaggleDBQA~\cite{kaggledbqa}, BIRD~\cite{DBLP:journals/corr/abs-2305-03111} etc. 
With these great efforts, the research communities have made impressive progress in \nlsql. 

Recently, large language models (LLMs), such as GPT-4 \cite{gpt4} from OpenAI and LLaMA \cite{llama} from Meta, have emerged as a milestone for natural language processing and machine learning. 
% brief intro to LLMs
Different from general machine learning model, LLMs are pre-trained on massive text corpus, which can perform various natural language tasks.
% how to work
The basic operating principle is to gradually produce the next word that has the highest probability based on the input prompt~\cite{DBLP:journals/corr/abs-2303-18223}. 
Therefore, to tackle \nlsql task with LLMs, the core is to find the optimal prompt, also known as prompt engineering~\cite{enhancing, DBLP:journals/corr/abs-2303-13547}. 

Specifically, according to number of examples provided in prompt, prompt engineering are classified into two scenarios: zero-shot scenario and few-shot scenario. 
In zero-shot scenario, no example is provided, and the main challenge is to represent the natural language question effectively, including incorporating relevant information such as the corresponding database schema~\cite{DBLP:journals/corr/abs-2305-11853, DBLP:journals/corr/abs-2303-13547, trummer2022codexdb, dong2023c3}. 
\revision{In this paper, the process of representing natural language questions and relevant information is referred to as {\bf{question representation}}.}
\revision{
While in few-shot scenario, a limited number of examples are available, thus besides question representation, we also need to study how to select the most helpful examples and organize them in the prompt appropriately. 
In natural language processing, the above progress that LLMs learn from contextual examples is named as {\bf{in-context learning}}~\cite{icl22}. 
It enables LLMs to identify explicit or inexplicit patterns from the input prompt, and generate corresponding outputs. 
In this way, LLMs are capable of new tasks during inference without any explicit task-specific training phase. 
Recent studies~\cite{din-sql, DBLP:journals/corr/abs-2304-13301, DBLP:conf/acl-deelio/LiuSZDCC22} confirm the significant role of including examples for effective in-context learning.
}
In this paper, we will discuss in-context learning in the scope of example selection and example organization.

\revision{Although LLMs are demonstrated to be effective in both zero-shot and few-shot scenarios in prior studies~\cite{DBLP:journals/corr/abs-2305-01598, enhancing, DBLP:journals/corr/abs-2303-13547, sun2023sql, DBLP:journals/corr/abs-2305-11853}, their performances can be further enhanced by {\bf{supervised fine-tuning (SFT)}}, which enhances LLMs using additional task-specific training data to make it more suitable for specific downstream tasks. 
In recent researches, supervised fine-tuning is used as a training paradigms of \textbf{Alignment}, which aligns LLMs' behavior to avoid generating offensive, biased responses and hallucinations~\cite{chatgpt}. 
In this paper, we will focus on enhancing LLMs' \nlsql capabilities with supervised fine-tuning.
}
It is worth noting that despite the extensive research on prompt engineering for \nlsql, there is a scarcity of studies exploring the supervised fine-tuning of LLMs for \nlsql~\cite{sun2023sql}, leaving this area as an open question. 

In summary, question representation, in-context learning, together with supervised fine-tuning are three essential knobs in large language model based \nlsql. 
In this paper, we will provide a systematical study and discussion about them. 

\section{Methodology}

As stated above, in this paper we focus on question representation, in-context learning and supervised fine-tuning. 
In this section, we provide formal definitions for these three problems, survey their existing solutions systematically, and point out the potential issues in existing techniques. 
To address these issues, we propose a new \nlsql prompt engineering method, named \ours, which refreshes the best performance in Spider leaderboard with $86.6\%$ execution accuracy.

\subsection{Question Representation}
\label{subsec:question_representation}

\begin{table}[t]
\small
	\begin{tabular}{cccccccc}
		\toprule
		\makecell{Question \\Representation}	& INS & RI  & FK  &	Ref.	& LLMs   &	\makecell{EX \\(\%)}	\\
		\hline
		\abbsprompt	& \ding{55} & \ding{55} & \ding{55} &	\cite{din-sql} & -   & -	\\ \hline
		\abtextprompt	& \ding{51} & \ding{55} & \ding{55} &	\cite{enhancing}	& CODE-DAVINCI-002   & 69.0	\\ \hline
            \multirow{2}{*}{\abopenaiprompt}	& \multirow{2}{*}{\ding{51}} & \multirow{2}{*}{\ding{51}} & \multirow{2}{*}{\ding{55}} &	\cite{DBLP:journals/corr/abs-2303-13547}	& GPT-3.5-TURBO  & 70.1 \\
		                  & & & &	\cite{din-sql}	& GPT-4 &	64.9 \\ \hline
		\multirow{3}{*}{\absqlprompt}	& \multirow{3}{*}{\ding{51}} & \multirow{3}{*}{\ding{55}} & \multirow{3}{*}{\ding{51}} & \cite{enhancing} & CODE-DAVINCI-002 &	75.6  \\
                        & & & & \cite{DBLP:journals/corr/abs-2305-11853} & CODE-DAVINCI-002 &	71.8  \\
                        & & & & \cite{DBLP:journals/corr/abs-2305-11853} & GPT-3.5-TURBO &	70.7  \\ \hline
		\abalpacaprompt	& \ding{51} & \ding{55} & \ding{55} &	\cite{alpaca}	& -	& - 	\\
		\bottomrule
	\end{tabular}
	\caption{Question representations in existing works, as well as their reported execution accuracy (EX) in zero-shot scenario. The Instruction (INS), Rule Implication (RI) and Foreign Key (FK) are possible components in a prompt. INS is the task description, such as ``Write a SQL to answer the question''. RI is the guiding statement, such as ``Complete sqlite SQL query only and with no explanation''. FK is the foreign key information of the database.} 
    \label{tab:related_work}
\end{table}

In this section, we first discuss question representations under zero-shot scenario for \nlsql.
Considering a target question $q$ in natural language on certain database $\mathcal{D}$, the target of question representation is to maximize the possibility of LLM $\mathcal{M}$ generating the correct SQL $s^*$ as follows:
\begin{align*} 
    \centering
    \underset{\sigma}{\max}\qquad &\mathbb{P}_{\mathcal{M}}(s^*|\sigma(q, \mathcal{D})),
\end{align*}
where function $\sigma(\cdot, \cdot)$ decides representation for target question $q$, with the useful information from the schema of database $\mathcal{D}$. 
Besides, $\sigma(\cdot, \cdot)$ also can include information such as instruction statement, rule implication and foreign key.
{
\begin{lstlisting}[language=Prompt, caption={Example of \bsprompt}, float=t, label={lst:bsprompt}]
Table continents, columns = [ContId, Continent]
Table countries, columns = [CountryId, CountryName, Continent]
Q: How many continents are there?
A: SELECT
\end{lstlisting}

\begin{lstlisting}[language=Prompt, caption={Example of \textprompt}, label={lst:textprompt}, float=t]
Given the following database schema:
continents: ContId, Continent
countries: CountryId, CountryName, Continent

Answer the following: How many continents are there?
SELECT
\end{lstlisting}
}

Follow the above definition, we survey different choices of $\sigma$ in zero-shot scenario and choose four most representative ones from literature. 
In addition, we also include the question representation used in Alpaca~\cite{alpaca} since it's popular in supervised fine-tuning. 
\tabref{tab:related_work} summarizes these five representation methods and lists their reported details from their original papers.

\begin{itemize}
    \item \textbf{\bsprompt} (\abbsprompt). 
    \bsprompt~\cite{din-sql} is a simple representation shown in \lstref{lst:bsprompt}. 
    It is consisted of table schemas, natural language question prefixed by ``\textit{Q: }'' and a response prefix ``\textit{A: SELECT}'' to prompt LLM to generate SQL. 
    In this paper we named it as \bsprompt due to its absence of instructions.
    
    \item \textbf{\textprompt} (\abtextprompt). 
    As shown in \lstref{lst:textprompt}, \textprompt \cite{enhancing} represents both schema and question in natural language.
    Compared with \bsprompt, it adds instruction at the very beginning of prompt to guide LLMs. In \cite{enhancing}, it achieves $69.0\%$ execution accuracy on Spider-dev in zero-shot scenario. 

    \item \textbf{\openaiprompt} (\abopenaiprompt). 
    The \openaiprompt (\lstref{lst:openaiprompt}) is first used in OpenAI's official \nlsql demo \cite{openaiprompt}, and evaluated in \cite{DBLP:journals/corr/abs-2303-13547, din-sql}. 
    It's consisted of instruction, table schemas, and question, where all information are commented by pound sign ``$\textit{\#}$''. 
    Compared with \textprompt, the instruction in \openaiprompt is more specific with a rule, ``\textit{Complete sqlite SQL query only and with no explanation}'', which we will further discuss in the \secref{sec:chatgpt} along with experimental results. 
    
    \item \textbf{\sqlprompt} (\absqlprompt). 
    The \sqlprompt \cite{DBLP:journals/corr/abs-2305-11853, enhancing} presents \nlsql task in SQL syntax. 
    Specifically, as shown in \lstref{lst:sqlprompt}, it directly presents ``\textit{CREAT TABLE}'' SQLs, and prompts LLM with natural language question in comments. 
    Compared with other representations, \absqlprompt stands out due to its ability to provide comprehensive information necessary for database creation, such as column types and primary/foreign keys.
    With such a representation, \cite{enhancing} correctly predicts about $75.6\%$ SQLs with LLM CODE-DAVINCI-002. 
    
    \item \textbf{\alpacaprompt} (\abalpacaprompt). 
    The \alpacaprompt is a prompt designed for supervised fine-tuning \cite{alpaca}. 
    As shown in \lstref{lst:alpacaprompt}, it prompts LLM to follow instruction and finish task according to the input context in Markdown format. 
    We include it to examine its effectiveness and efficiency in both prompt engineering and supervised fine-tuning scenarios. 
    
\end{itemize}

\begin{lstlisting}[language=Prompt, caption={Example of \openaiprompt}, label={lst:openaiprompt}, float=t]
### Complete sqlite SQL query only and with no explanation
### SQLite SQL tables, with their properties:
# 
# continents(ContId, Continent)
# countries(CountryId, CountryName, Continent)
# 
### How many continents are there?
SELECT
\end{lstlisting}

\begin{lstlisting}[language=Prompt, caption={Example of \sqlprompt}, label={lst:sqlprompt}, float=t]
/* Given the following database schema: */
CREATE TABLE continents(
	ContId int primary key,
    Continent text,
	foreign key(ContId) references countries(Continent)
);

CREATE TABLE countries(
	CountryId int primary key,
    CountryName text,
    Continent int,
	foreign key(Continent) references continents(ContId)
);

/* Answer the following: How many continents are there? */
SELECT 
\end{lstlisting}

\begin{lstlisting}[language=Prompt, caption={Example of \alpacaprompt}, label={lst:alpacaprompt}, float=t
]
Below is an instruction that describes a task, paired with an input that provides further context. Write a response that appropriately completes the request.

### Instruction:
Write a sql to answer the question "How many continents are there?"

### Input:
continents(ContId, Continent)
countries(CountryId, CountryName, Continent)

### Response:
SELECT 
\end{lstlisting}

As shown in \tabref{tab:related_work}, different representations are experimented with different LLMs, and integrated in different frameworks, making it difficult to compare them fairly and effectively. 
Additionally, the specific roles played by individual components such as foreign key information and rule implication remain unclear. 
Consequently, it is essential to conduct a systematical study to better understand question representations, and investigate their advantages and disadvantages through a fair comparison.

\subsection{In-Context Learning for \nlsql}
The above question representation methods enable LLMs to directly output desired SQLs by zero-shot learning. 
However, LLMs can perform better for \nlsql through in-context learning, in which only a few examples are provided in the input prompts. Therefore, in this subsection, we discuss the keys of in-context learning, that are example selection and example organization. We first give a formulation of in-context learning to ease the further discussions. 

In \nlsql, given a set of triples $\mathcal{Q} = \{(q_i, s_i, \mathcal{D}_i)\}$, where $q_i$ and $s_i$ are natural language question and its corresponding SQL query on database $\mathcal{D}_i$, the target of in-context learning for \nlsql is to maximize the possibility of LLM $\mathcal{M}$ generating the correct SQL $s^*$ on the target question $q$ and database $\mathcal{D}$ as follows:
\begin{align*} 
    \underset{\mathcal{Q}', \sigma}{\max}\qquad &\mathbb{P}_{\mathcal{M}}(s^*|\sigma(q, \mathcal{D}, \mathcal{Q}')), \\
    \mathrm{s.t.}\qquad|&\mathcal{Q}'| = k\quad\mathrm{and} \quad\mathcal{Q}' \subset \mathcal{Q},
\end{align*}
where function $\sigma(\cdot, \cdot, \cdot)$ decides representation for target question $q$, with the useful information from the schema in database $\mathcal{D}$ and $k$ examples selected from $\mathcal{Q}$. In this paper, we focus on \emph{cross-domain \nlsql}, which means the target database $\mathcal{D}$ is not included among the databases $\mathcal{D}_i$ mentioned in $\mathcal{Q}$., i.e., $\mathcal{D} \notin \{\mathcal{D}_i|(q_i, s_i, \mathcal{D}_i) \in \mathcal{Q}\}$.

In-context learning for \nlsql involves selecting the most helpful examples $\mathcal{Q'}$ and deciding how to organize the information of these selected examples into prompt. 
Next we discuss these two sub-tasks: example selection and example organization.

\subsubsection{Example Selection}
\label{subsubsec:exp_select}

We summarize various example selection strategies in prior studies as follows.

\begin{itemize}
    \item \textbf{Random}. This strategy randomly samples $k$ examples from the available candidates. Previous works~\cite{DBLP:conf/acl-deelio/LiuSZDCC22, DBLP:journals/corr/abs-2304-13301, enhancing} have adopted it as a baseline for example selection. 
    \item \textbf{\qsselector} (\abqsselector). \abqsselector~\cite{DBLP:conf/acl-deelio/LiuSZDCC22} chooses $k$ examples with the most similar questions. 
    Specifically, it embeds both example questions in $\mathcal{Q}$ and the target question $q$ with a pre-trained language model. 
    Then it applies a pre-defined distance measure, such as the Euclidean distance or negative cosine similarity, to each example-target pair. Finally $k$NN algorithm is leveraged to select $k$ examples from $\mathcal{Q}$ that closely match the target question $q$. 

    \item \textbf{\slmselector} (\abslmselector). For cross-domain \nlsql, \abslmselector~\cite{DBLP:journals/corr/abs-2304-13301} eliminates the negative influence of domain-specific information by replacing table names, column names, and values in all questions with a mask token, and then compute the similarities of their embedding with $k$NN algorithm.
    \item \textbf{\qrsselector} (\abqrsselector). Instead of using the target question $q$, 
    \abqrsselector~\cite{enhancing} aims to select $k$ examples that are similar to target SQL query $s^*$. 
    Specifically, it employs a preliminary model to generate SQL query $s'$ using target question $q$ and database $D$, where this generated $s'$ can be regarded as an approximation of $s^*$. 
    Then it encodes queries from examples into binary discrete syntax vectors according to their keywords. 
    After that, it chooses $k$ examples by considering both similarity to the approximated query $s'$ and diversity among selected examples. 
\end{itemize}

Above strategies focus on selecting examples using only target question or query. 
However, according to prior studies~\cite{icl22}, in-context learning is essentially learning from analogy. 
In the case of \nlsql, the objective is to generate queries that match the given questions, thus LLMs are supposed to learn the mapping from questions to SQL queries. 
Therefore, we point out that during example selection, taking both question and SQL queries into consideration may benefit \nlsql task. 
We will further discuss it in \secref{sec:dail-sql}.

\subsubsection{Example Organization} 
The example organization plays a pivotal role in determining what information of the above selected examples will be organized into the prompt. 
We summarize existing strategies in prior studies into two categories, \fiorg and \sqlorg, 
as demonstrated in \lstref{lst:org_complete} and \lstref{lst:org_sql}. 
% In our study, we explore two existing strategies for example organization, as demonstrated in \lstref{lst:org_complete} and \lstref{lst:org_sql}. 
In these examples, $\textit{\$\{DATABASE\_SCHEMA\}}$ represents the database schema, and $\textit{\$\{TARGET\_QUESTION\}}$ stands for the question representation in \lstref{lst:sqlprompt}.

\begin{lstlisting}[language=Prompt, label={lst:org_complete}, caption={Example of \fiorg.}, float=t]
/* Given the following database schema: */
${DATABASE_SCHEMA}
/* Answer the following: How many authors are there? */
SELECT count(*) FROM authors

/* Given the following database schema: */
${DATABASE_SCHEMA}
/* Answer the following: How many farms are there? */
SELECT count(*) FROM farm

${TARGET_QUESTION}
\end{lstlisting}

\begin{lstlisting}[language=Prompt, label={lst:org_sql}, caption={Example of \sqlorg.}, float=t]
/* Some SQL examples are provided based on similar problems: */
SELECT count(*) FROM authors

SELECT count(*) FROM farm

${TARGET_QUESTION}
\end{lstlisting}

\begin{lstlisting}[language=Prompt, label={lst:org_qa}, caption={Example of \pairorg.}, float=t]
/* Some example questions and corresponding SQL queries are provided based on similar problems: */
/* Answer the following: How many authors are there? */
SELECT count(*) FROM authors

/* Answer the following: How many farms are there?. */
SELECT count(*) FROM farm

${TARGET_QUESTION}
\end{lstlisting}

\begin{itemize}
    \item \textbf{\fiorg} (\abfiorg). 
    \abfiorg~\cite{enhancing, DBLP:journals/corr/abs-2305-11853} organizes examples in the same representation with the target question.  
    As shown in \lstref{lst:org_complete}, examples are structured identically to the target question, and the only difference is that instead of the ``\textit{SELECT}'' token at the end, the selected examples have the corresponding SQL queries after ``\textit{SELECT}''. 
    \item \textbf{\sqlorg} (\absqlorg).
    \absqlorg~\cite{DBLP:journals/corr/abs-2304-13301} includes only SQL queries of the selected examples with a prefix instruction in the prompt, as demonstrated in~\lstref{lst:org_sql}.
    Such organization aims at maximizing the number of examples with limited token length. 
    However, it removes the mapping information between questions and corresponding SQL queries, and such information can be useful, which we will demonstrate later.
\end{itemize}

In summary, \abfiorg includes the full information of examples, which ensures the quality; 
while \absqlorg only keeps SQL queries to accommodate more examples, which prefers the quantity. 
We wonder if there exists a better trade-off between quality and quantity in example organization
, which can further benefit the \nlsql task.

\subsection{\ours}
\label{sec:dail-sql}
To address the aforementioned issues in example selection and organization, in this subsection, we present a novel \nlsql method named \ours. 
\revision{Please refer to \appref{dail-sql:pseudocode} for the pseudocode of \ours.}

For example selection, inspired by \abslmselector and \abqrsselector, we proposed \textbf{\pqsselector} (\abpqsselector), considering both questions and queries to select candidates. 
Specifically, \pqsselector first masks domain-specific words in both target question $q$ and example questions $q_i$ in the candidate set $\mathcal{Q}$. 
It then ranks the candidate examples based on the Euclidean distance between the embeddings of masked $q$ and $q_i$. Simultaneously, it calculates the query similarity between the pre-predicted SQL query $s'$ and $s_i$ in $\mathcal{Q}$. 
Finally, the selection criterion prioritizes the sorted candidates by question similarity with a query similarity greater than a predefined threshold $\tau$. In this way, the selected top $k$ examples have good similarity with both question and query.

To preserve the mapping information between questions and SQL queries and also improve the token efficiency, we propose a new example organization strategy \textbf{\pairorg} (\abpairorg) to trade-off in terms of quality and quantity.
Specifically, \abpairorg presents both questions $q_i$ and corresponding SQL queries $s_i$, as illustrated in~\lstref{lst:org_qa}. As a compromise between \abfiorg and \absqlorg, \abpairorg reserves the question-SQL mapping, and reduces the token length of examples by removing token-cost database schema.

In \ours, we adopt \absqlprompt as our question representation.
The reason is that compared with other representations, \absqlprompt contains full information of the database, including primary and foreign keys, which may offers more useful information for LLMs, such as foreign keys for the prediction of ``JOIN'' clauses. 
Besides, pre-trained on extensive coding corpora, LLMs could better understand the prompt in \absqlprompt without too much additional effort.

In summary, \ours utilizes \absqlprompt as the question representation, selects examples based on information from both question and query, and organizes them to keep question-to-SQL mappings. In such prompt design, LLMs could work better for \nlsql task, and in Spider leaderboard, the proposed \ours refresh the performance with $86.2\%$ execution accuracy.

Note \ours is a flexible LLM-based \nlsql solution, which can be further extended and integrated with other components easily. 
For example, to improve the performance, we equip \ours with self-consistency~\cite{DBLP:conf/iclr/0002WSLCNCZ23}, which achieves a performance of $86.6\%$ execution accuracy. 
Although self-consistency improves the execution accuracy by $0.4\%$, it is very time consuming and yields many times the cost of original \ours. 
Therefore, in this paper we still focus on \ours.

\subsection{Supervised Fine-Tuning for \nlsql}
To enhance the performance of LLMs in zero-shot scenario, the popular option for existing \nlsql methods is in-context learning, which is \revision{discussed} in above subsections. As an alternative yet promising option, supervised fine-tuning is less explored so far.  
Similar to supervised fine-tuning for various language task, we can adopt it to the field of \nlsql, and improve LLMs' performance on this downstream task. 
To further understand how supervised fine-tuning works for \nlsql, we first provide a brief formulation as follows.

For \nlsql, given a large language model $\mathcal{M}$, a set of \nlsql training data $\mathcal{T}=\{(q_i, s_i, \mathcal{D}_i)\}$, where $q_i$ and $s_i$ are the natural language question and its corresponding query on database $\mathcal{D}_i$, the objective of supervised fine-tuning is to minimize the following empirical loss:
\begin{equation*}
    \min_{\sigma, \mathcal{M}^*} \sum_{i=1}^{|\mathcal{T}|}{\mathcal{L}_{\mathcal{M^*}}(\sigma(q_i, \mathcal{D}_i), s_i)},
\end{equation*}
where $\mathcal{L}$ is the loss function to measure the difference between the generated query and the groundtruh query. Similar to question representation, $\sigma$ decides question representation with useful information from the schema in database $\mathcal{D}$. 
In this definition, supervised fine-tuning for \nlsql covers two sub-tasks, including fine-tuning the given LLM $\mathcal{M}$ using supervised data $\mathcal{T}$ in order to get the optimal LLM $\mathcal{M}^*$, and searching for the optimal question representation $\sigma$. 
Since question representations have been discussed in~\secref{subsec:question_representation}, this section will primarily focus on data preparation $\mathcal{T}$ and fine-tuning. 

For general domain, each item in supervised data $\mathcal{T} = \{(p_i, r_i)\}$ contains an input prompt $p_i$ and an expected respond $r_i$ from LLM. 
To ensure consistency with the inference process, we employ a supervised fine-tuning and generate prompt-response pairs from a given \nlsql dataset. Specifically, given a \nlsql data set $\mathcal{T} = \{(q_i, s_i, \mathcal{D}_i)\}$, we fine-tune the LLMs using the generated tuning data by using target question and the given database as prompt, and treating the desired query as response from LLM, i.e.,  $\mathcal{T} = \{(p_i = \sigma(q_i, \mathcal{D}_i), r_i = s_i)\}$. 
Once the data is ready, we can use existing package to fine-tune the given LLM $\mathcal{M}$ through either full fine-tuning~\cite{instructgpt} or parameter-efficient fine-tuning~\cite{DBLP:conf/iclr/HuSWALWWC22} depending on the available computational resources. After fine-tuning, the optimized LLM $\mathcal{M}^*$ can be used to do inference, that is asking it to generate queries through natural language questions.
Note that we utilize the same question representation $\sigma$ in both fine-tuning and inference processes.
We will conduct a series of experiments and discuss the great potential of supervised fine-tuning for \nlsql.
\begin{figure*}[t]
    \centering
    \includegraphics[width=0.48\linewidth]{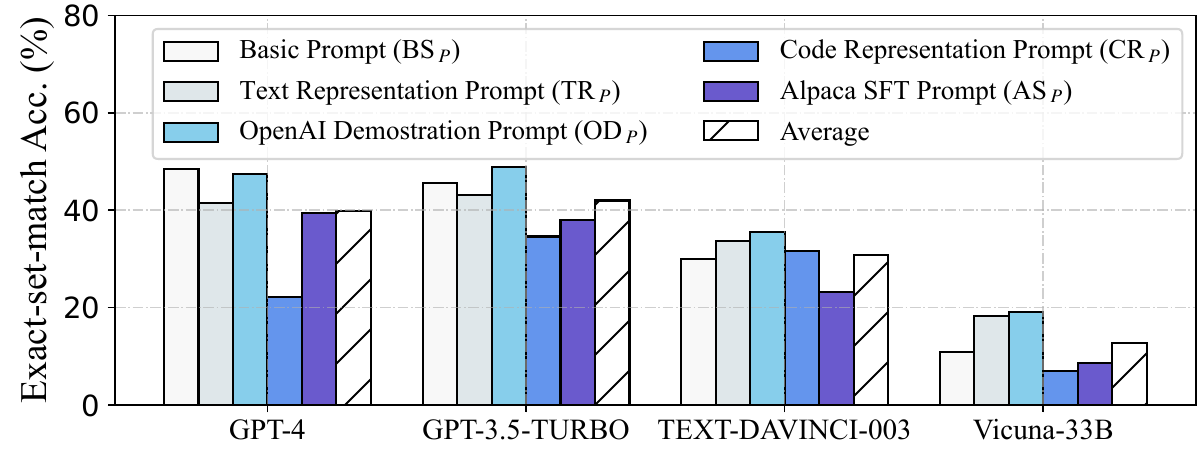}
    % \hfill
    \includegraphics[width=0.48\linewidth]{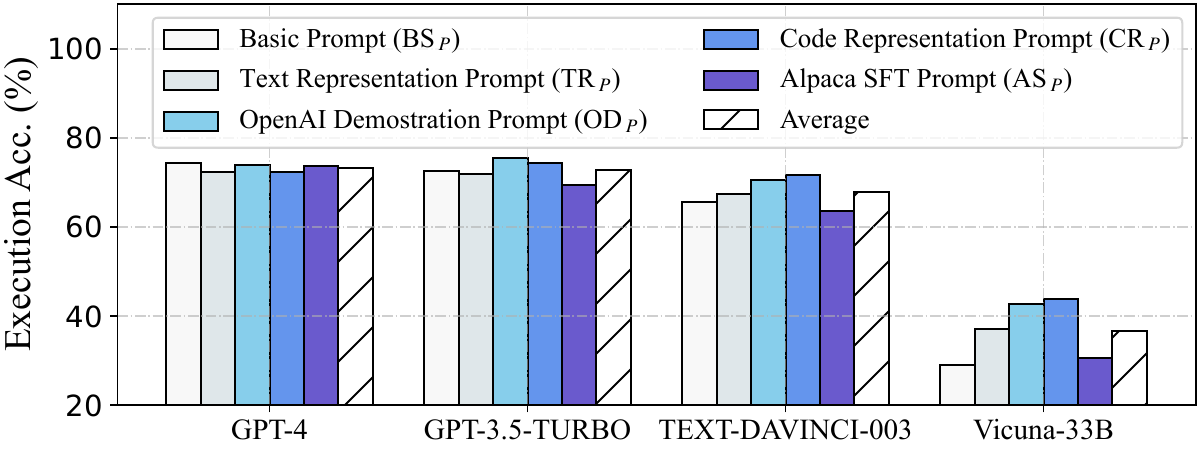}
    \caption{\revision{Results of different question representations on Spider-dev under zero-shot scenario.}}
	\label{fig:openai_zero_shot}
\end{figure*}

\begin{figure*}[t]
    \centering

    \includegraphics[width=0.48\linewidth]{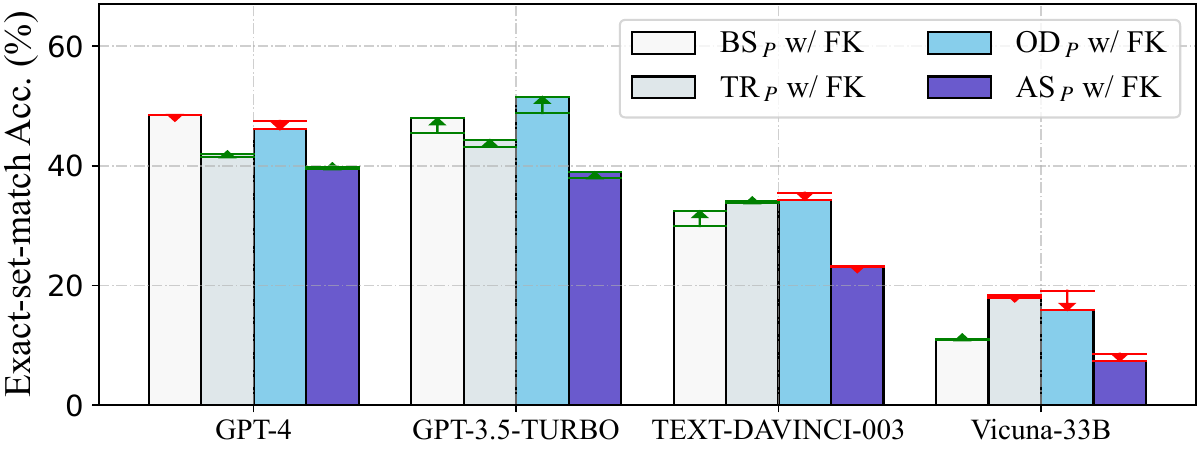}
    \includegraphics[width=0.48\linewidth]{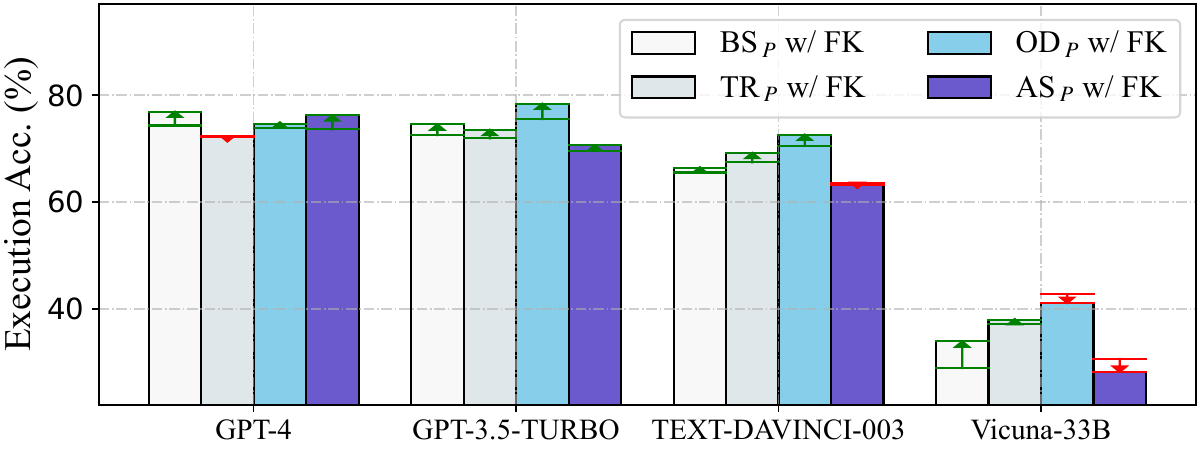}
    \caption{\revision{Ablation studies of foreign keys information on Spider-dev. The \textcolor{green}{green} arrow indicates an increase, and \textcolor{red}{red} arrow indicates a decrease.}}
	\label{fig:openai_zero_shot_key}
\end{figure*}

\begin{figure*}[t]
    \centering

    \includegraphics[width=0.48\linewidth]{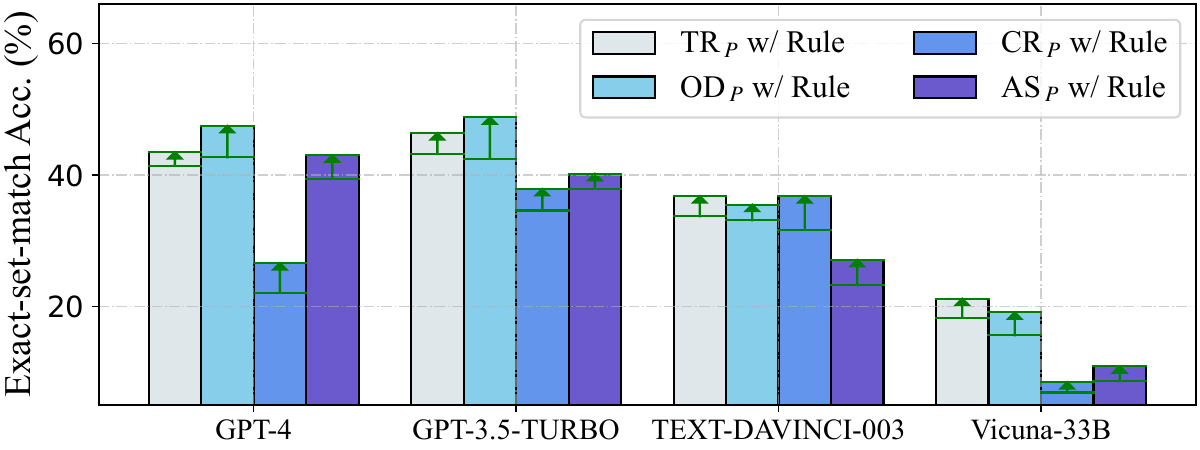}
    \includegraphics[width=0.48\linewidth]{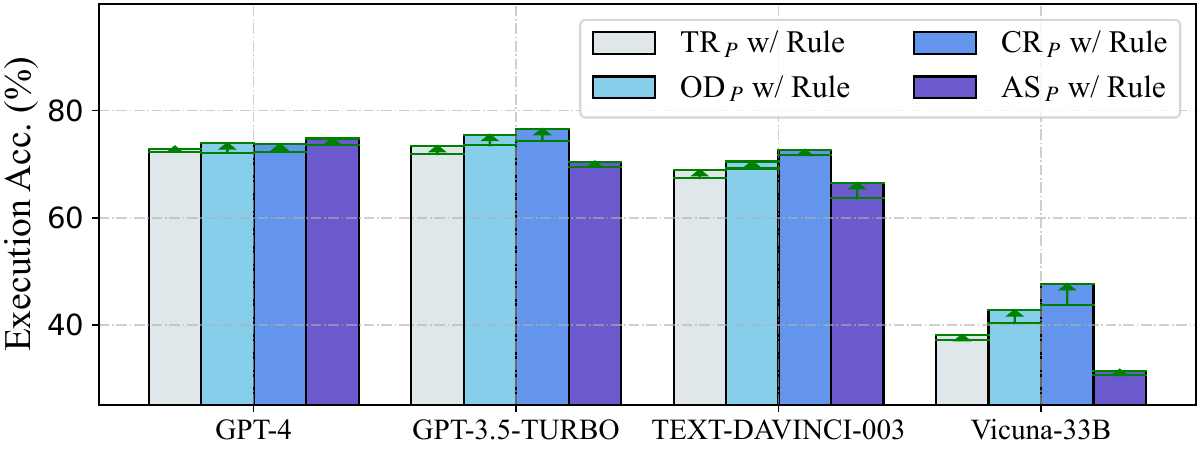}
    \caption{\revision{Ablation studies of ``with no explanation'' rule implication on Spider-dev. The \textcolor{green}{green} arrow indicates an increase, and \textcolor{red}{red} arrow indicates a decrease.}}
	\label{fig:openai_zero_shot_rule}
\end{figure*}

\section{Experiment}
\label{sec:exp}
In this section, we first introduce our experimental settings. 
Then we conduct extensive comparisons with existing solutions in question representation, in-context learning and supervised fine-tuning respectively. 
After that, we further compare them in terms of token efficiency to inspire more efficient solutions.

\subsection{Setting}

\textbf{Dataset}. 
We evaluate \nlsql methods on two well recognized datasets, \textbf{Spider} \cite{spider} and \textbf{Spider-Realistic} \cite{realistic}. 
Spider is a large-scale cross-domain \nlsql dataset, which contains 8659 instances in training split and 1034 instances in development split over 200 databases. 
Each instance is consisted of a natural language question on a specific database and its corresponding SQL query. 
In this paper, we use the development split \emph{Spider-dev} for the purpose of evaluation as the test split is not released.
Spider-Realistic \cite{realistic} is a more challenging variant of Spider. 
It selects a subset of 508 examples from Spider-dev and manually revises the questions while keeping the SQL queries unchanged. 
For few-shot scenarios, we utilize the training split of Spider as the example candidates when testing with both Spider-dev and Spider-Realistic.

\textbf{Metric}. 
To make a fair comparison, we follow prior study~\cite{DBLP:conf/emnlp/ZhongYK20} to use exact-set-match accuracy ({\bf{EM}}) and execution accuracy ({\bf{EX}}). 
The exact-set-match accuracy measures the matched SQL keywords between the predicted SQL query and its corresponding ground truth. 
The execution accuracy, on the other hand, compares the execution output of the predicted SQL query with that of the ground truth SQL query on some database instances. 
This metric provides a more precise estimate of the model's performance since there may be multiple valid SQL queries for a single given question.
We use the existing released evaluation scripts at~\href{https://github.com/taoyds/test-suite-sql-eval}{https://github.com/taoyds/test-suite-sql-eval}.

\textbf{LLM}.
To ensure a fair comparison, for all the methods, we use the same maximal context length, that is 4096 for OpenAI LLM and 2048 for open-source LLM. 
During evaluation, we leave 200 tokens for response generation. 
By default, we set the argument temperature as 0 to eliminate the influence of randomness. 
Regarding post-processing, we follow existing work to extract the first SQL query in response and remove additional output. 
For more implementation details, please refer to \appref{app:engineering}.

\begin{table*}[th]
        \small
	\centering
	\begin{tabular}{llcccccccccc}
		\toprule
		\multirow{2}{*}{Few-shot}	& \multirow{2}{*}{Selection}	& \multirow{2}{*}{\makecell{Question \\Similarity}}    & \multirow{2}{*}{\makecell{Query \\Similarity}}    &	\multicolumn{2}{c}{\gptfour} &	\multicolumn{2}{c}{\chatgpt}	&	\multicolumn{2}{c}{\davinci}	& \multicolumn{2}{c}{\revision{\vicuna}}\\
		\cmidrule(r){5-6}   \cmidrule(r){7-8}  \cmidrule(r){9-10} \cmidrule(r){11-12}
		&	 & & & EM	&	EX  & EM	&	EX	&	EM	&	EX	&	\revision{EM}	&	\revision{EX} \\	\hline
            0-shot  & -    &  - &  - & 22.1	& 72.3	& 34.6	& 74.4	& 31.7	& 71.7 & \revision{6.9}	& \revision{43.7}   \\  \hline
		\multirow{5}{*}{1-shot}	&	\abrandselector	& 0.23   & 0.47 & 41.7  & 77.4   & 45.9 &	73.9 & 38.2   & 70.6		& \revision{14.4}	& \revision{47.9}   \\
		&	Question Similarity selection	& 0.39  & 0.65 & 53.3  & 78.8   & 51.9  & 74.3	 & 44.1   &	72.3			& \revision{16.5}	& \revision{48.5}   \\
		&	Masked Question Similarity selection	& 0.57  & 0.80 & 58.2 & 79.1  & 57.4  & 76.0	 & 47.9   &	75.0			& \revision{21.4}	& \revision{48.7}   \\
		&	DAIL selection	& 0.56  & 0.95 & 62.1   & 80.2   & 59.5  & 75.5   & 51.9  & 76.9				& \revision{22.8}	& \revision{49.2}   \\
            &	Upper Limit	& 0.56 & 0.98 & 63.7  & 81.0   & 61.4  & 77.2  & 53.1  & 77.5			& \revision{22.7}	& \revision{49.4}   \\\hline
		\multirow{5}{*}{3-shot}	&	\abrandselector	& 0.23  & 0.48  & 48.9   & 79.4  & 49.0   & 73.6  & 41.7  &	71.6	& \revision{16.8}	& \revision{46.9}   \\
		&	Question Similarity selection	& 0.37  & 0.63 & 56.3  & 79.2  & 53.8  & 74.7  & 52.2  & 74.1				& \revision{21.1}	& \revision{47.1}   \\
		&	Masked Question Similarity selection	& 0.54 & 0.78  & 66.1   & 81.5  & 61.1   & 77.3  & 59.7  &	77.0			& \revision{27.7}	& \revision{52.3}   \\
		&	DAIL selection	& 0.53 & 0.94  & 69.1  & 81.7  & 63.9   & 77.8  & 64.4  & 79.5			& \revision{30.7}	& \revision{53.6}   \\
            &	Upper Limit	& 0.53 & 0.98  & 71.5   & 83.4  & 66.2   & 79.2  & 66.7  &	81.1			& \revision{31.2}	& \revision{54.4}   \\\hline
		\multirow{5}{*}{5-shot}	&	\abrandselector	& 0.23  & 0.48  & 51.6  & 79.5  & 52.9  & 75.7  & 49.0  & 72.1			& \revision{-}	& \revision{-}   \\
		&	Question Similarity selection	& 0.36  & 0.61 & 58.2  & 79.9  & 55.9  & 75.1  & 54.8  & 73.2				& \revision{-}	& \revision{-}   \\
		&	Masked Question Similarity selection	& 0.52 & 0.77 & 66.8  & 82.0  & 62.3  & 77.9  & 64.7  & 78.6				& \revision{-}	& \revision{-}   \\
		&	DAIL selection	& 0.52 & 0.94 & 71.9 & 82.4  & 66.7  & 78.1  & 67.7  & 80.5				& \revision{-}	& \revision{-}   \\
            &	Upper Limit	& 0.51 & 0.97 & 74.4  & 84.4  & 68.8  & 79.6  & 70.7  & 82.4			& \revision{-}	& \revision{-}   \\%\hline
		\bottomrule
	\end{tabular}
	% \caption{Results of different selection strategies on Spider-dev with few-shot evaluation.}
        \caption{\revision{Evaluation on Spider-dev with different example selections. The organization is fixed to \fiorg.}}
	\label{tab:kk_prompt_spider}
\end{table*}

\subsection{Question Representations}
In this subsection, we evaluate the question representations presented in~\secref{subsec:question_representation} under zero-shot scenario,  \revision{employing four LLMs: \gptfour, \chatgpt, \davinci, and \vicuna}.

\figref{fig:openai_zero_shot} presents the comparison of different question representations over Spider-dev. 
By comparing different representations, we can observe that \abopenaiprompt fits to all \revision{four} LLMs and achieves $75.5\%$ execution accuracy with GPT-3.5-TURBO. 
In contrast, \abalpacaprompt exhibits poor performance with \revision{\chatgpt, \davinci, and \vicuna}, necessitating a suitable LLM to work well with.
Unexpectedly, \gptfour exhibits a preference for the simple \abbsprompt derived from Din-SQL~\cite{din-sql}, indicating that a powerful LLM can mitigate the complexities associated with representation design.
Besides, by comparing the average performance for \revision{four} LLMs, \gptfour and \chatgpt are more capable in the zero-shot scenario. 
Due to the expensive cost of \gptfour, \chatgpt together with \abopenaiprompt maybe a better choice for the zero-shot scenario. \revision{For less powerful LLMs like \davinci and \vicuna, \abopenaiprompt and \absqlprompt are preferred.}
For detailed numerical results, please refer to \appref{app:prompt_spider}. 

To further investigate the different question representations, we conduct ablation study to explore the effects of their invidual components. 

\textbf{Foreign Key} (FK). 
Foreign Key implies the relation among different relational tables, which might be helpful in \nlsql task. 
In our evaluation, only \absqlprompt contains foreign key information. 
To examine its effect, we add foreign key information into other representations and evaluate them in~\figref{fig:openai_zero_shot_key}. 
\revision{For OpenAI LLMs, we observe that foreign key significantly improves the execution accuracy of LLMs by $0.6\%-2.9\%$, except the combinations of \abtextprompt with \gptfour ($-0.2\%$) and \abalpacaprompt with \davinci ($-0.4\%$).} \revision{However, the impact of foreign key for \vicuna tends to be unstable. 
Notably, the inclusion of foreign keys leads to a surprising improvement of $5.0\%$ for the \abbsprompt, but adversely affects the performance of the \abopenaiprompt and \abalpacaprompt.}

\textbf{Rule Implication} (RI). 
Inspired by the outperformance of \abopenaiprompt, we explore the effect of rule implication. 
Specifically, \abopenaiprompt implicate LLMs to generate SQL queries ``\textit{with no explanation}''
To examine the effect of ``\textit{with no explanation}'' rule in question representation, we present an ablation study in \figref{fig:openai_zero_shot_rule}. 
% Specifically, we remove ``\textit{with no explanation}'' from \abopenaiprompt, and add it to other representations.
\revision{Specifically, we plot the performance of different representations after including the ``\textit{with no explanation}" implication and the change of accuracy.}
From~\figref{fig:openai_zero_shot_rule} we observe adding this rule consistently booms the performance of all LLMs in both exact-set-match and execution accuracy, with the most significant improvements exceeding $6\%$ and \revision{$3\%$}, respectively. 
While for \abopenaiprompt, removing this rule incurs about $2.4\%-6.2\%$ drop in exact-set-match accuracy, and \revision{$1.3\%-2.4\%$} drop in execution accuracy, indicating the importance of this rule implication. 
As a comparison, we also test a popular rule implication ``\textit{Let's think step by step}''~\cite{kojima2022large}, which guides LLM to generate response with analysis.
However, its performance is highly volatile in \nlsql task as \appref{app:step_by_step} shows. 
Due to limited resources, we leave the exploration of other possible rule implications as an open question for future research. 

In summary, both the foreign key and the ``\textit{with no explanation}'' implication rule are beneficial for \nlsql task.
In our evaluation, \abopenaiprompt with foreign keys and GPT-3.5-TURBO are the most effective and economic combination, which achieves $51.5\%$ exact-set-match accuracy and $78.4\%$ execution accuracy. 
\subsection{In-Context Learning for \nlsql}
\label{sec:chatgpt}

In few-shot scenario, we examine different example selection and organization strategies with \revision{\gptfour, \chatgpt, \davinci, and \vicuna}. 
To ensure a fair comparison, we adopt \absqlprompt as the question representation for all the experiments in this subsection, due to its superior performance in one-shot preliminary experiment in~\appref{app:prompt_spider_1shot}.

\begin{figure*}[th]
    \centering
    \begin{minipage}[t]{0.98\linewidth}
    \subfigure[\revision{\gptfour on Spider-dev}]{\includegraphics[width=0.25\linewidth]{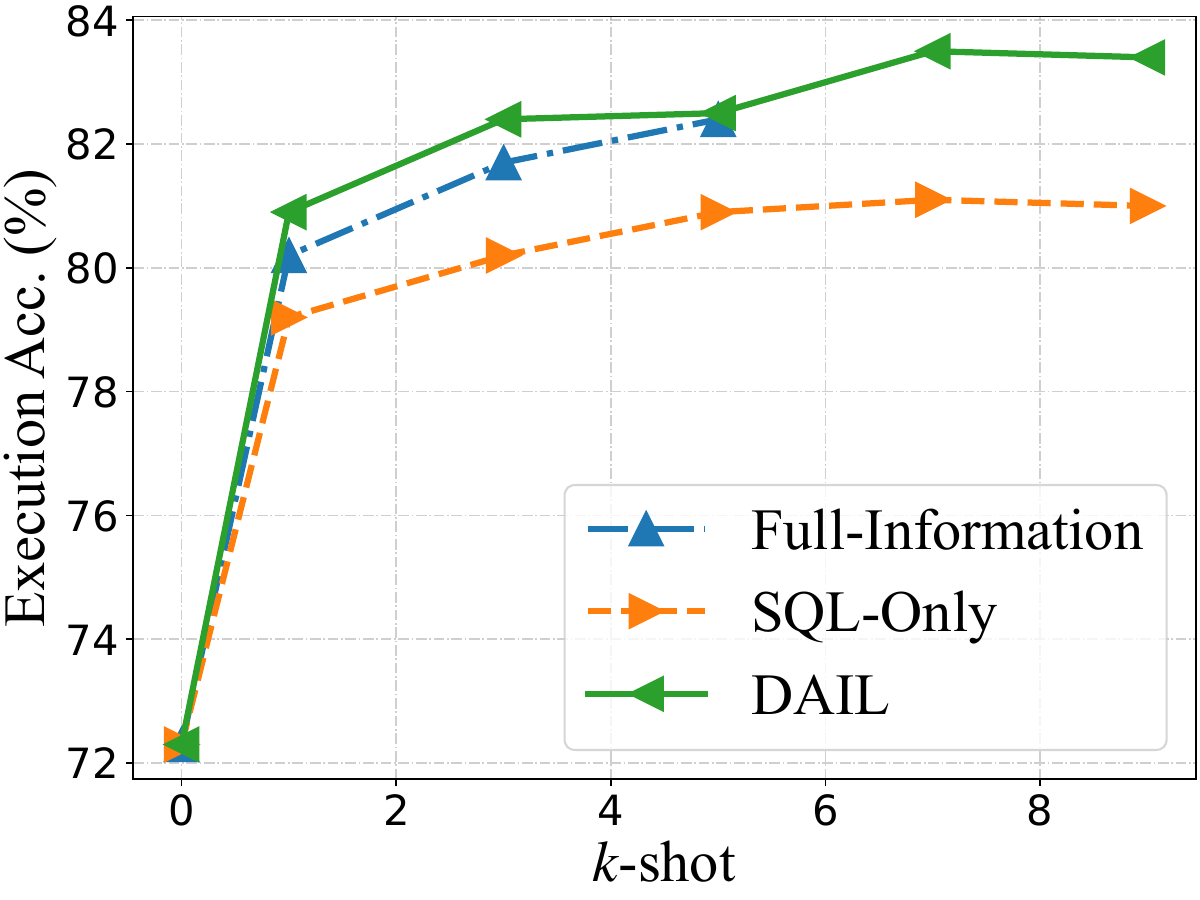}\label{fg:organization:dev:gpt4}}%
    \hfil%
    \subfigure[\revision{\chatgpt on Spider-dev}]{\includegraphics[width=0.25\linewidth]{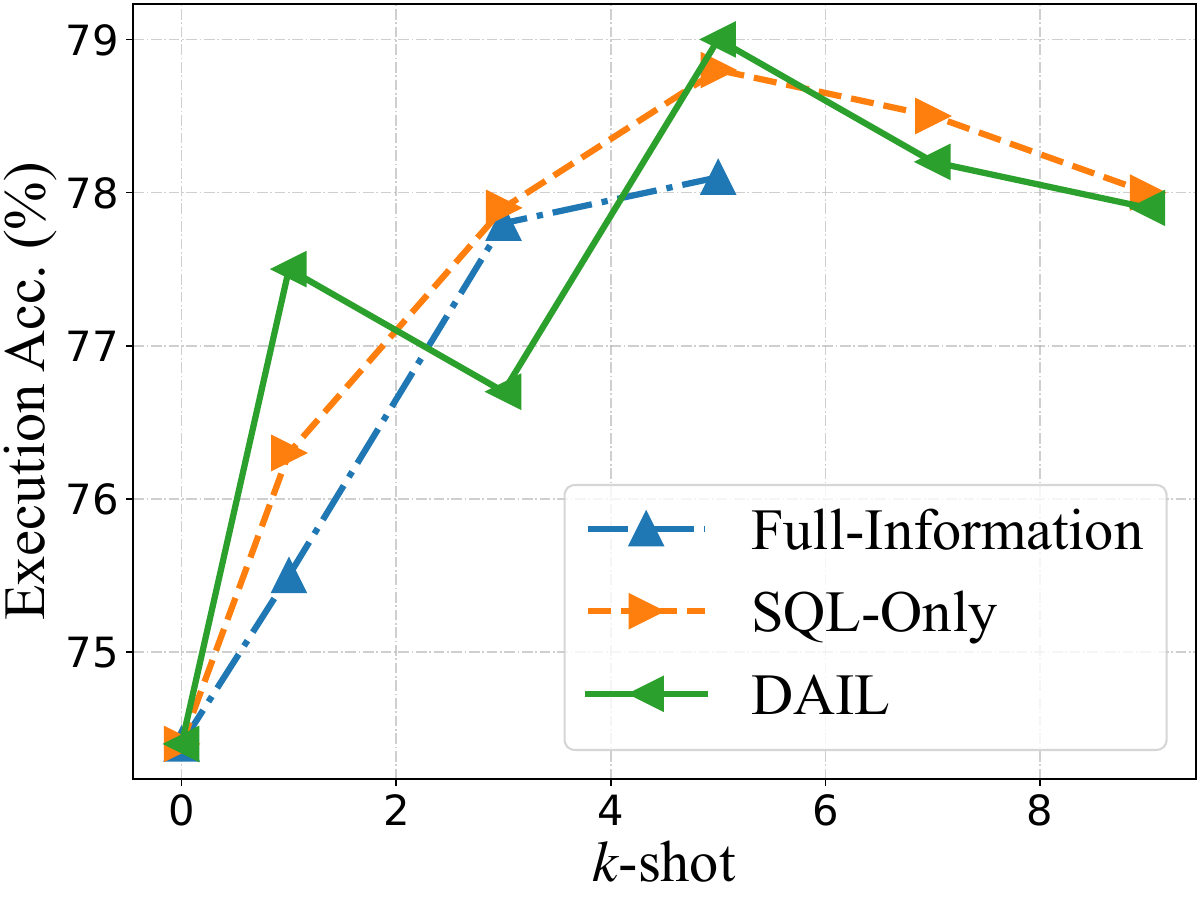}\label{fg:organization:dev:chatgpt}}%
    \hfil%
    \subfigure[\revision{\davinci on Spider-dev}]{\includegraphics[width=0.25\linewidth]{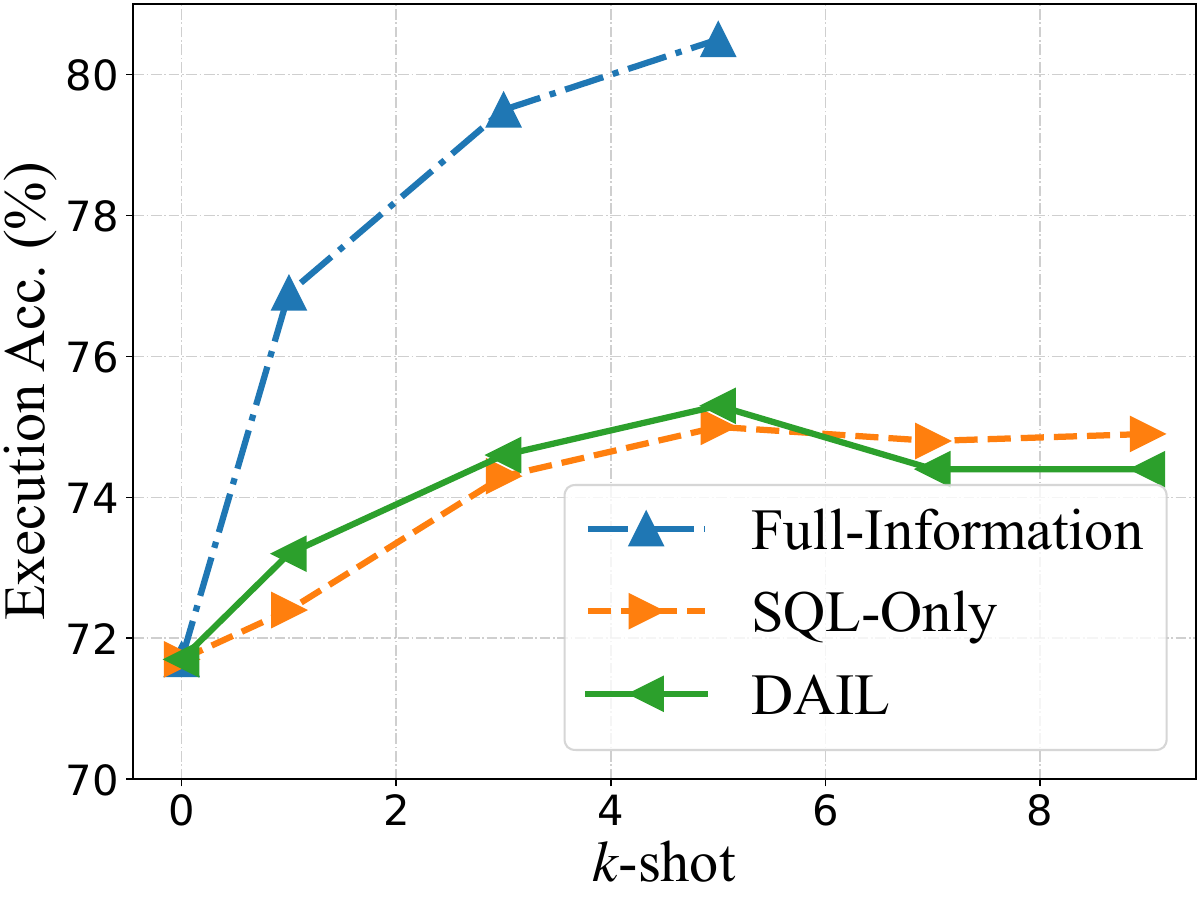}\label{fg:organization:dev:davinci}}%
    \hfil%
    \subfigure[\revision{\vicuna on Spider-dev}]{\includegraphics[width=0.25\linewidth]{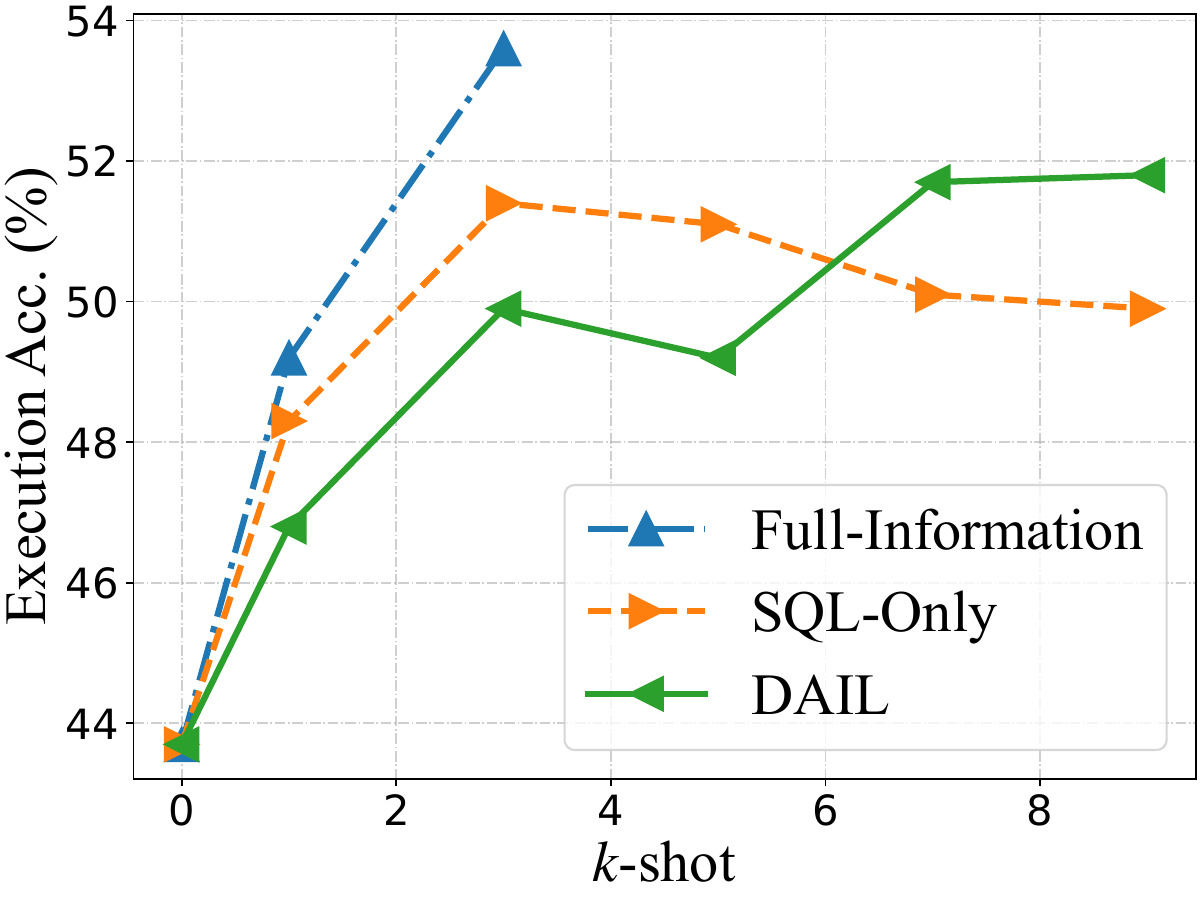}\label{fg:organization:dev:vicuna}}%
    \end{minipage}
    \begin{minipage}[t]{0.98\linewidth}
    \subfigure[\revision{\gptfour on Spider-Realistic}]{\includegraphics[width=0.25\linewidth]{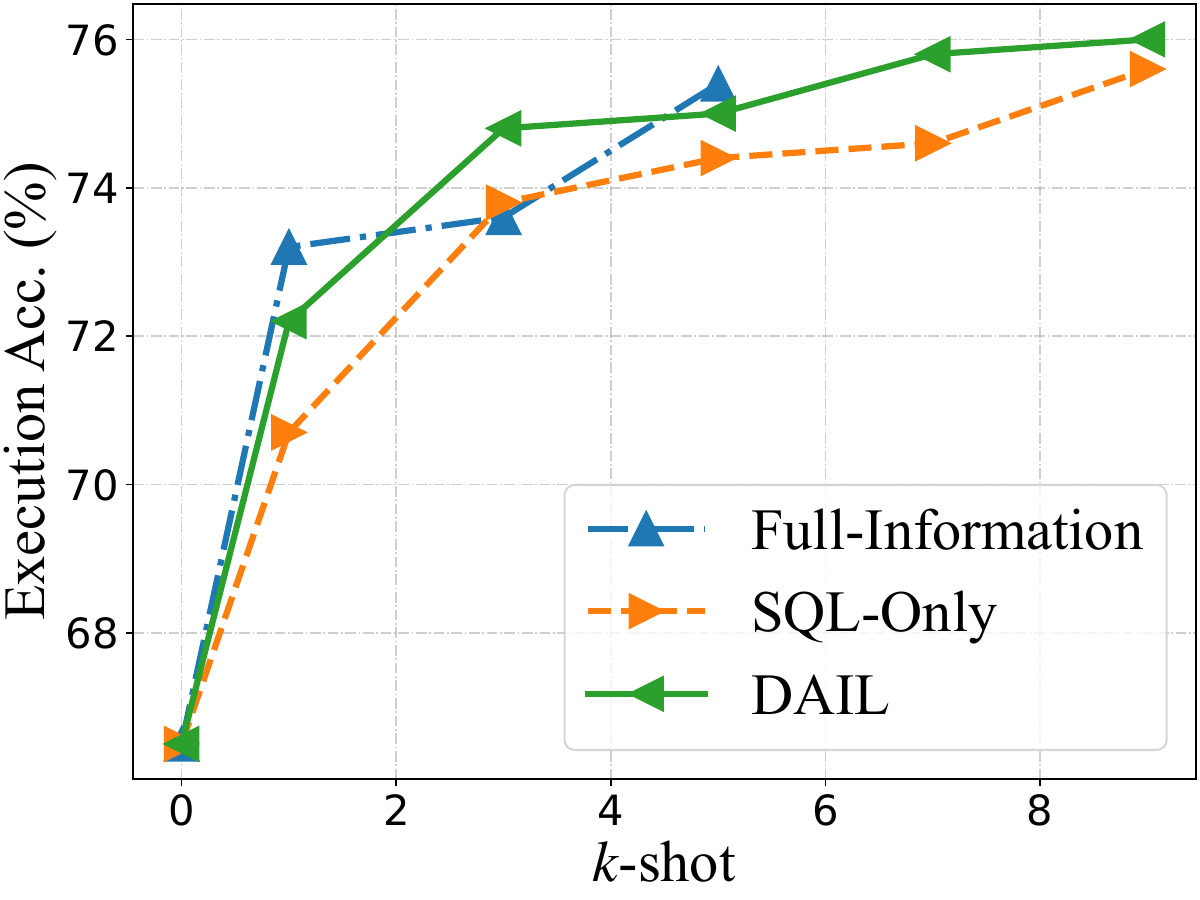}\label{fg:organization:realistic:gpt4}}%
    \hfil%
    \subfigure[\revision{\chatgpt on Spider-Realistic}]{\includegraphics[width=0.25\linewidth]{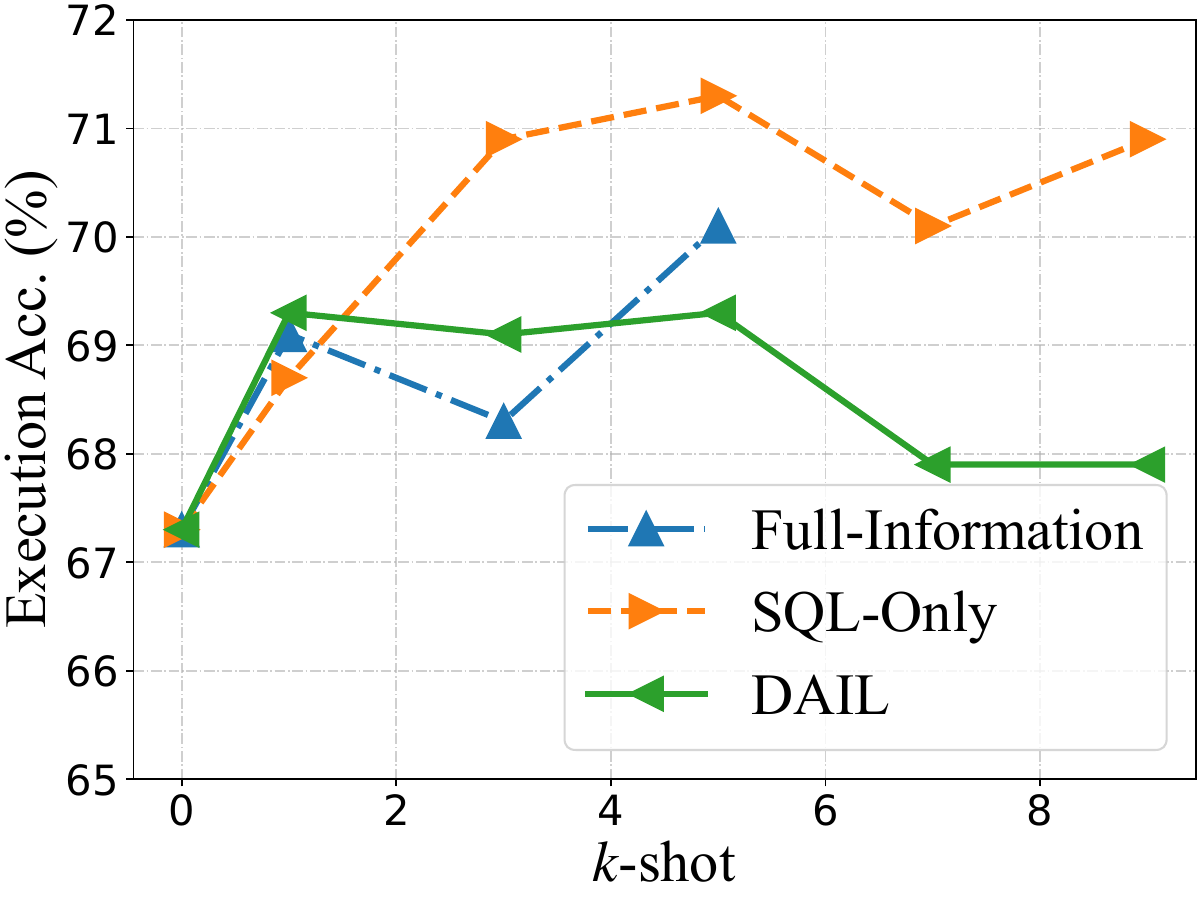}\label{fg:organization:realistic:chatgpt}}%
    \hfil%
    \subfigure[\revision{\davinci on Spider-Realistic}]{\includegraphics[width=0.25\linewidth]{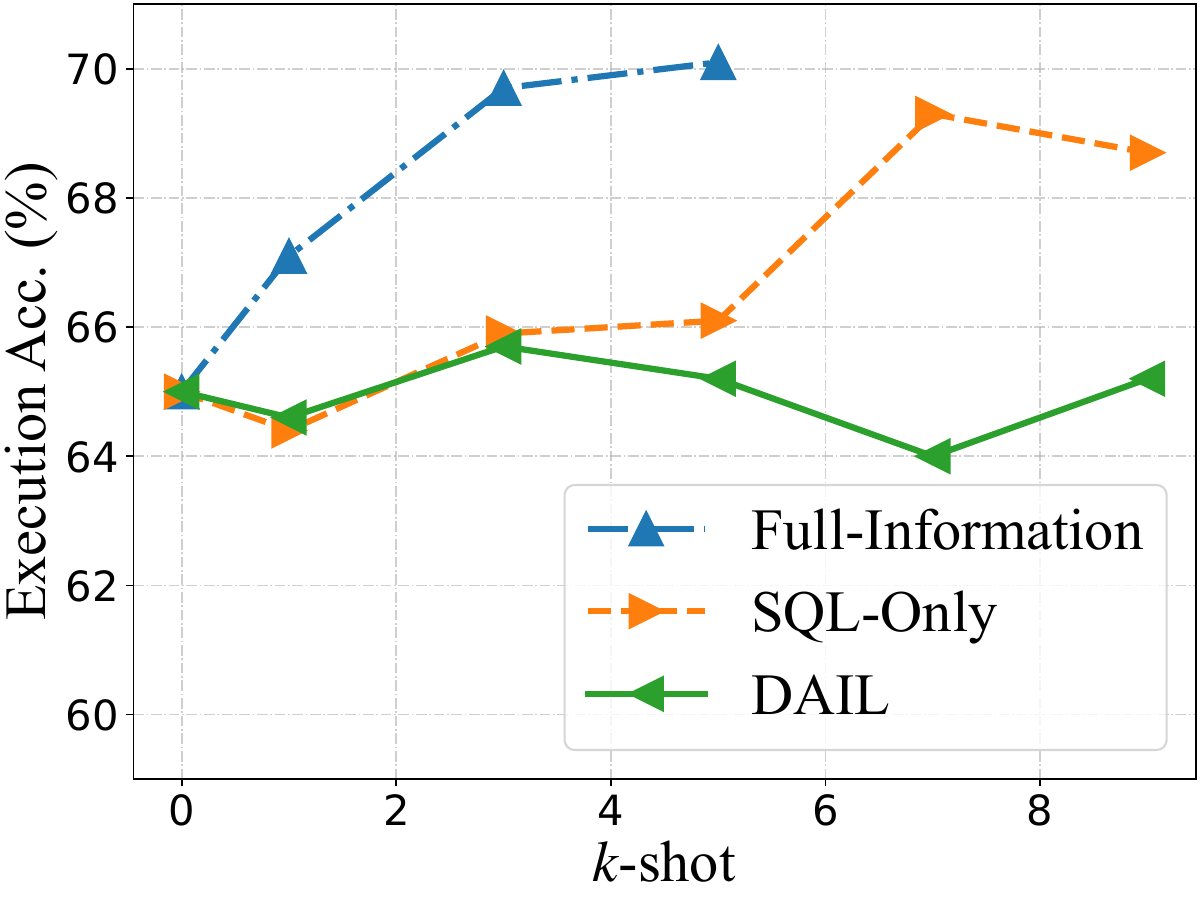}\label{fg:organization:realistic:davinci}}%
    \hfil%
    \subfigure[\revision{\vicuna on Spider-Realistic}]{\includegraphics[width=0.25\linewidth]{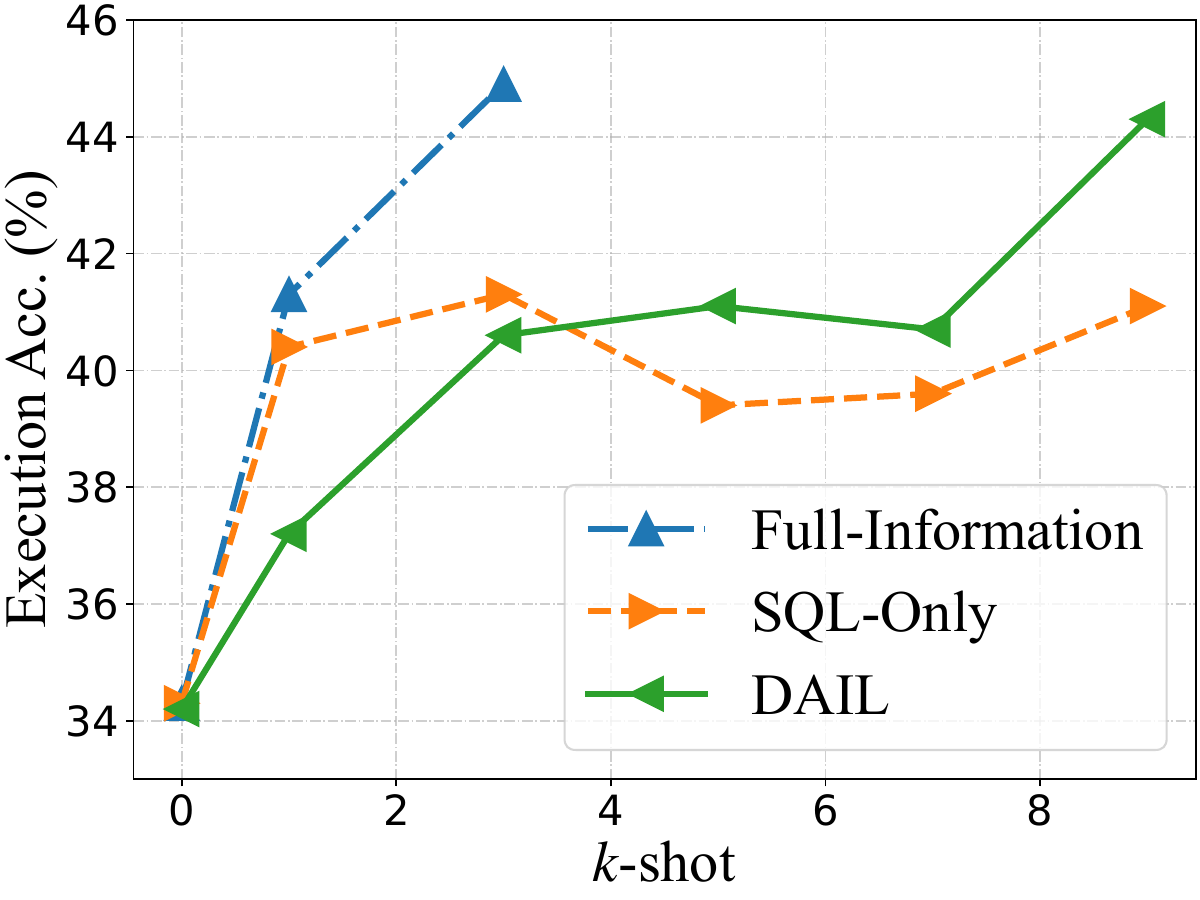}\label{fg:organization:realistic:vicuna}}%
    \caption{\revision{Evaluation on Spider-dev with different example organizations. The selection is fixed to DAIL Selection.}}\label{fg:organization}
    \end{minipage}
\end{figure*}

\subsubsection{Example Selection}
To verify the importance of both question and query for example selection, we calculate question's and query's Jaccard similarities between chosen examples and the target instance, and report the averaged numbers under column \emph{question similarity} and \emph{query similarity} in~\tabref{tab:kk_prompt_spider}. 
Specifically, we remove database-specific information from questions~\cite{rat-sql} and queries~\cite{li2023resdsql}, and calculate the Jaccard similaritis of the remained tokens. 
\revision{Besides, we introduce \textbf{Upper Limit} for reference, which is similar with \pqsselector but utilizes the ground truth query $s^*$ rather than the query generated by preliminary predictor. Notably, we do not directly provide the ground truth SQL to the LLMs, but just use the ground truth query as a reference for selecting examples.}
To some extent, Upper Limit indicates the upper bound of performance for similarity based selection methods.

\tabref{tab:kk_prompt_spider} shows the comparisons of different example selection strategies in 1-, 3- and 5-shot scenarios on Spider-dev.
By comparing different selection strategies, it is demonstrated that \abpqsselector generally outperforms other strategies. 
In 5-shot scenario, equipped with GPT-4, \ours achieves $82.4\%$ execution accuracy. 
Besides, in \tabref{tab:kk_prompt_spider} we observe the increasing question and query similarity corresponds to higher execution accuracy mostly, indicating the importance of considering both question and query similarity. 
Note \abpqsselector's execution accuracy is still lower than Upper Limit. 
This discrepancy can be attributed to the lower query similarity, indicating the gap between the ground truth query and that generated by the preliminary model.

\subsubsection{Example Organization}

To compare different example organization strategies, we evaluate \fiorg, \sqlorg and \pairorg in few-shot scenario on both Spider-dev and Spider-Realistic. 
\figref{fg:organization} shows the comparison results, and refer to \appref{app:exp_organization} for detailed numerical results. 

From \figref{fg:organization:dev:gpt4} and \figref{fg:organization:realistic:gpt4}, we can observe that \gptfour benefits from contextual examples steadily on both Spider-dev and Spider-Realistic. 
With \pairorg, its execution accuracy increases from $72.3\%$ to $83.5\%$ on Spider-dev and from $66.5\%$ to $76.0\%$ on Spider-Realistic. 
While for \chatgpt and \davinci, adding examples may incur drop in execution accuracy due to limited in-context learning capability. 
\revision{Regarding \vicuna, its performance consistently improves as the number of examples increases in \pairorg.}
By comparing different organization strategies, we observe that \gptfour shows preference for \pairorg in both Spider-dev and Spider-Realistic, suggesting it can effectively learn the mapping from question-SQL pairs. 
For \chatgpt (\figref{fg:organization:dev:chatgpt} and \figref{fg:organization:realistic:chatgpt}), compared with its zero-shot performance in \figref{fig:openai_zero_shot}, its enhancement in in-context learning is the smallest among \revision{four} LLMs, due to its weakness in in-context learning.
For \davinci, \fiorg is far beyond the other two strategies, especially with increasing example number, as depicted in~\figref{fg:organization:dev:davinci} and \figref{fg:organization:realistic:davinci}. 
\revision{Figures \ref{fg:organization:dev:vicuna} and \ref{fg:organization:realistic:vicuna} illustrate that in the case of Vicuna-33B, \pairorg outperforms \sqlorg but falls short of the performance achieved by \fiorg.}
By comparing different LLMs, we infer that for LLM with greater in-context learning capability, like GPT-4, benefits from \pairorg the most, while the weaker LLMs require more information to learn from examples. 
However, we emphasize \pairorg can be a good choice to achieve higher performance, and 
the best execution accuracy in our evaluation is achieved by \pairorg with GPT-4.

In summary, for example selection, our findings emphasize the importance of the mapping from question to SQL query. 
Considering both question and query similarities simultaneously, \abpqsselector outperforms other selection strategies in our evaluation. 
For example organization, we show the effectiveness of \abpairorg, and point out its demands for potent LLMs. 
Finally, in our evaluation, we observe that our approach, \ours, equipped with GPT-4, achieves the highest performance with an execution accuracy of $83.5\%$ on Spider-dev and $76.0\%$ on Spider-Realistic. \revision{For more comparisons with previous methods, please refer to \appref{app:exp_cmp_plm_rule}.}

\begin{table*}[t]
	\centering
	\begin{tabular}{llcc cc cc cc cc cc}
		\toprule
		\multirow{2}{*}{}	& 	\multirow{2}{*}{LLM}	&	\multicolumn{2}{c}{\abbsprompt}	&	\multicolumn{2}{c}{\abtextprompt}	&	\multicolumn{2}{c}{\abopenaiprompt}	&	\multicolumn{2}{c}{\absqlprompt}	&	\multicolumn{2}{c}{\abalpacaprompt}	&	\multicolumn{2}{c}{Average}	\\
        \cmidrule(r){3-4} \cmidrule(r){5-6} \cmidrule(r){7-8} \cmidrule(r){9-10} \cmidrule(r){11-12} \cmidrule{13-14}
		&				&	EM	&	EX	&	EM	&	EX	&	EM	&	EX	&	EM	&	EX	&	EM	&	EX	&	EM   &	EX    \\	\hline
        % \hline
        \multirow{3}{*}{Pre-trained}	&	LLaMA-7B	&	6.5	&	9.6	&	3.1	&	4.9	&	3.6	&	9.0	&	4.8	&	16.3	&	1.3	&	5.9	&	3.9	&	9.1	\\
	&	LLaMA-13B	&	8.8	&	18.4	&	4.5	&	15.2	&	8.2	&	21.8	&	5.6	&	25.0	&	8.9	&	26.9	&	7.2	&	21.5	\\
	&	LLaMA-33B	&	9.6	&	26.7	&	12.0	&	25.9	&	13.6	&	36.4	&	12.2	&	\textbf{42.8}	&	\textbf{13.8}	&	38.1	&	12.2	&	34.0	\\
        &	\revision{Falcon-40B}	&	\revision{0.3}	&	\revision{11.7}	&	\revision{0.2}	&	\revision{0.9}	&	\revision{0.3}	&	\revision{7.6}	&	\revision{0.1}	&	\revision{21.9}	&	\revision{0.0}	&	\revision{5.0}	&	\revision{0.2}	&	\revision{9.4}	\\
		\hline																									
    \multirow{9}{*}{Aligned}	&	\delete{Alpaca-7B}	&	\delete{15.1}	&	\delete{25.1}	&	\delete{13.5}	&	\delete{23.8}	&	\delete{14.7}	&	\delete{25.7}	&	\delete{16.0}	&	\delete{32.1}	&	\delete{8.9}	&	\delete{19.9}	&	\delete{13.6}	&	\delete{25.3}	\\
	&	\delete{GPT4ALL-7B}	&	\delete{7.8}	&	\delete{19.4}	&	\delete{8.8}	&	\delete{24.6}	&	\delete{8.1}	&	\delete{27.0}	&	\delete{8.5}	&	\delete{25.9}	&	\delete{6.5}	&	\delete{21.8}	&	\delete{7.9}	&	\delete{23.7}	\\
	&	Vicuna-7B	&	7.5	&	15.6	&	1.2	&	9.9	&	6.2	&	21.5	&	5.6	&	24.0	&	0.9	&	5.4	&	4.3	&	15.3	\\
	&	Vicuna-13B	&	8.2	&	21.7	&	10.1	&	24.4	&	11.2	&	31.4	&	5.8	&	33.5	&	4.7	&	20.0	&	8.0	&	26.2	\\
	&	Vicuna-33B	&	10.8	&	28.9	&	18.3	&	37.1	&	19.1	&	42.7	&	6.9	&	43.7	&	8.6	&	30.6	&	12.7	&	36.6	\\
 	&	LLaMA-2-CHAT-7B	&	14.3	&	23.4	&	7.2	&	15.5	&	6.3	&	12.3	&	12.2	&	25.5	&	5.0	&	20.5	&	9.0	&	19.4	\\
	&	LLaMA-2-CHAT-13B	&	18.8	&	32.6	&	15.4	&	30.5	&	11.1	&	22.3	&	20.7	&	40.0	&	16.9	&	36.2	&	16.6	&	32.3	\\
        &	\revision{LLaMA-2-CHAT-70B}	&	\revision{21.8}	&	\revision{46.2}	&	\revision{11.9}	&	\revision{33.9}	&	\revision{21.4}	&	\revision{45.5}	&	\revision{12.4}	&	\revision{44.0}	&	\revision{8.4}	&	\revision{28.6}	&	\revision{15.2}	&	\revision{39.6}	\\
        &	\revision{CodeLLaMA-34B}	&	\revision{\textbf{27.8}}	&	\revision{65.5}	&	\revision{15.9}	&	\revision{40.3}	&	\revision{25.8}	&	\revision{65.3}	&	\revision{24.3}	&	\revision{\textbf{68.5}}	&	\revision{22.4}	&	\revision{61.5}	&	\revision{23.2}	&	\revision{60.2}	\\
		\bottomrule
	\end{tabular}
 %\vspace{0.1in}
	\caption{Zero-shot evaluation results on Spider-dev with different open-source LLMs. The best performances of pre-trained and aligned LLM are in bold.}	
	\label{tab:0shot_spider_ss}
\end{table*}

\subsection{Supervised Fine-Tuning for \nlsql}
\label{sec:sft}

In this section, we investigate supervised fine-tuning in \nlsql. 
Due to the unaffordable cost of fine-tuning OpenAI LLMs, we focus on open-source LLMs. 
Given the fact that very few existing work adopt open-source LLMs and their performance remain unknown, we first undertake a thorough evaluation for open-source LLMs, employing various question representation, example selection and organization strategies. 
After that, we fine-tune open-source LLMs in \nlsql and observe their enhancement in both zero-shot and few-shot scenarios.

\subsubsection{Open-source LLM}

To investigate the potential of open-source LLM, we choose LLaMA~\cite{llama}, and its aligned variants in varying scales. 
They are detailed as follows. 
Note the aligned variants means the LLM is aligned to be more helpful, harmless and honest~\cite{hhh}, 
and the suffix "-7B" means the LLM has 7 billions parameters, the same meaning for "-13B" and "-33B".
\begin{itemize}
    \item \textbf{LLaMA-7B/13B/33B}~\cite{llama} is a collection of widely recognized open-source LLMs, which are pre-trained on massive corpus by Meta.
    \item \revision{\textbf{Falcon-40B}~\cite{penedo2023refinedweb} is pre-trained solely on massive corpus of refined web data.}
	\item \delete{\textbf{Alpaca-7B}~\cite{alpaca} is an aligned version of LLaMA-7B, which is fine-tuned with 52k instruction-following data generated by TEXT-DAVINCI-003.} 
	\item \delete{\textbf{GPT4ALL-7B}~\cite{gpt4all} is another aligned version of LLaMA-7B with about 800k data designed for helpful, harmless and honest AI assistant.} 
    \item \textbf{LLaMA-2-CHAT-7B/13B\revision{/70B}}~\cite{llama2} are up-to-date version of LLaMA. 
    They are both pre-trained and aligned, and outperform the previous version on most benchmarks. 
	\item \textbf{Vicuna-7/13/33B}~\cite{vicuna, vicuna2023} is a collection of open-source chatbot aligned from LLaMA with user-shared conversations. 
    Vicuna-13B~\cite{vicuna2023} is declared to perform similar to OpenAI ChatGPT and Google Bard, and outperforms LLaMA and Alpaca in most scenarios. 
        \item \revision{\textbf{CodeLLaMA-34B}~\cite{roziere2023code} is an aligned version of LLaMA-2-34B, which is fine-tuned with 500B tokens of code data.} 
\end{itemize}

\subsubsection{Zero-shot Scenario with Open-source LLM}

\tabref{tab:0shot_spider_ss} shows their zero-shot performances on Spider-dev with different question representations. 
Due to limited space, please refer to \appref{app:0shot_realistic_ss} for the performance on Spider-Realistic.
Next, we provide several analysis from aspects of question representations, model scale and alignment as follows. 

\textbf{Effect of Question Representation}. 
\revision{We can observe that the best performances is achieved by \absqlprompt with \revision{$68.5\%$} execution accuracy on Spider-dev. 
\delete{The possible reason is that full database knowledge in \absqlprompt (\sqlprompt) compensates the incapability of open-source LLMs.}
%After that, \abopenaiprompt is comparable to \absqlprompt on Spider-dev and outperforms \absqlprompt on Spider-Realistic by $2.7\%$ with Vicuna-33B. 
One possible reason is that \absqlprompt tends to stimulate the coding capability of LLMs. 
This effect is particularly evident in CodeLLaMA-34B, which only achieve $40.3\%$ execution accuracy with natural language-based \abtextprompt.}

% model scale
\textbf{Effect of Model Scale}.  
From the results we observe a positive correlation  between model scale and performance on \nlsql for both LLaMA and Vicuna. 
Specifically, the average execution match accuracy of LLaMA shows a notable progression from $9.1\%$ to $34.0\%$ on Spider-dev, and Vicuna shows a similar upward trend from $15.3\%$ to $36.6\%$. 
\revision{With the most parameter size, LLaMA-2-CHAT-70B improves the average performance to $39.6\%$.} 
In the more challenging dataset Spider-Realistic,
the same pattern can be observed and execution accuracy of LLaMA and Vicuna rise from  $7.56\%$ to $25.4\%$ and $12.3\%$ to $30.0\%$.

\textbf{Effect of Alignment}. 
From the results we observe that LLM alignment can benefit \nlsql. 
Specifically, with the same model scale, Vicuna outperforms LLaMA by about $5\%$ in execution accuracy on both Spider-dev and Spider-Realistic. 
\revision{For Falcon-40B, it performs poorly with all representations, attributable to the absence of dedicated code data in its training dataset. As a comparison, with carefully collected code data in the alignment stage, CodeLLaMA-34B exhibits a significant improvement in \nlsql task with similar model scale. Note that, CodeLLaMA-34B also outperforms LLaMA-2-CHAT-70B by an average of $20.6\%$ despite having only half the parameter of LLaMA-2-CHAT-70B. This highlights the crucial importance of the training corpus in LLMs.}

\revision{In conclusion, having more parameters in LLMs may hold certain potential benefits to \nlsql, but the training corpus (e.g., having task-specific training data) plays a more crucial role.}

\subsubsection{Few-shot Scenario with Open-source LLM}

\begin{figure}[t]
    \centering
    \includegraphics[width=\linewidth]{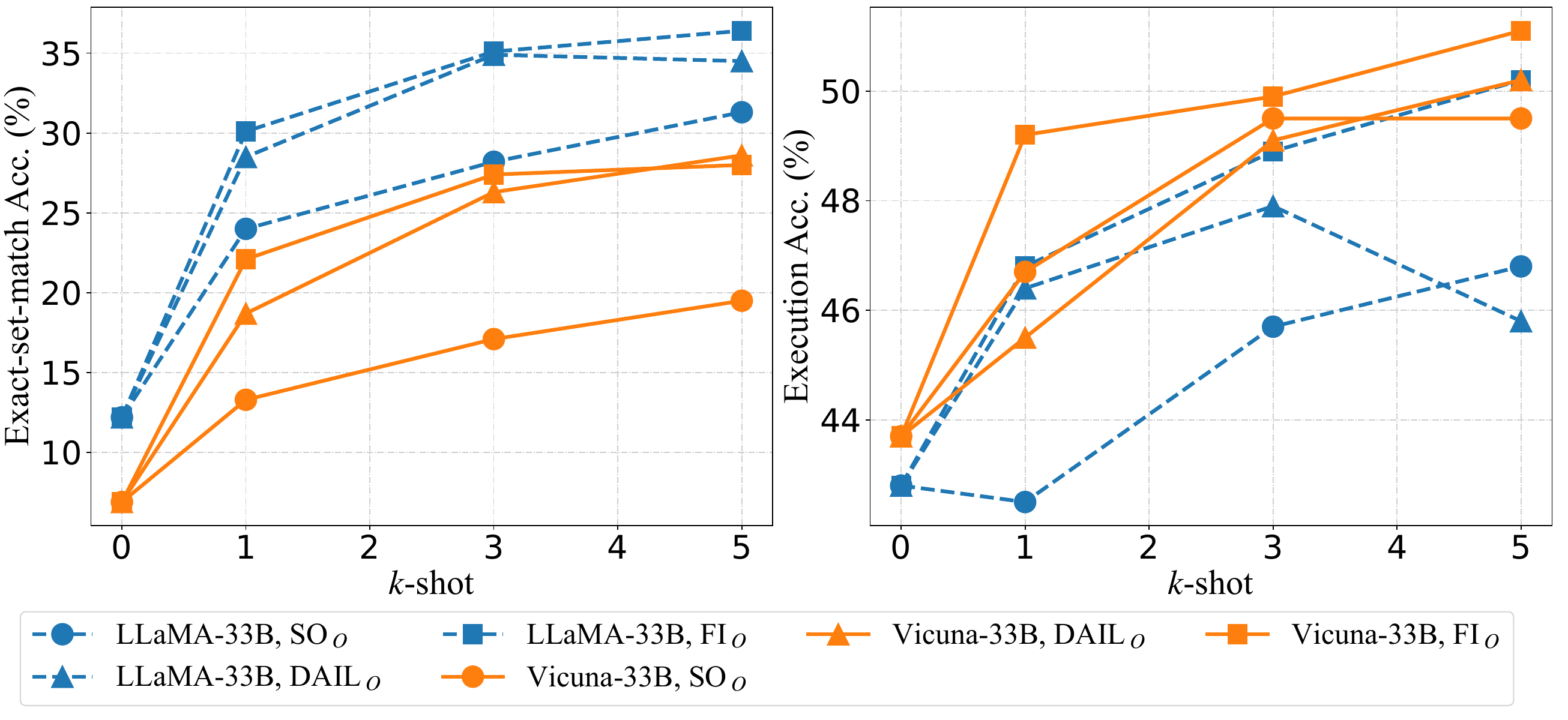}
    \caption{Few-shot evaluation with open-source LLMs on Spider-dev.}
	\label{fig:kshot_llms}
\end{figure}

For few-shot scenario, \figref{fig:kshot_llms} shows the performance of LLaMA-33B and Vicuna-33B with \absqlprompt.
We use \pqsselector to select example as it is reported as the best strategy in \secref{sec:chatgpt}. 
For more details, refer to \appref{app:kshot_llms}. 
From this Figure, we can see that LLaMA-33B benefits more than Vicuna-33B, and achieves $36.4\%$ exact-set-match accuracy with 5-shot \fiorg examples. 
Regarding execution match accuracy, increasing number of examples benefits \nlsql in most cases. 
Besides, among different organization strategies, \fiorg outperforms other strategies in different k-shot scenarios, which achieves $51.5\%$ execution accuracy with Vicuna-33B. 
\revision{Please refer to 
\appref{app:kshot_llms_size_and_corpus}
for more analysis in few-shot scenario.}

Notably, in both zero-shot and few-shot scenarios, the open-source LLMs are far behind OpenAI LLMs. 
We will try to further enhance their performance with supervised fine-tuning.

\begin{figure*}[th]
    \centering
    \includegraphics[width=0.48\linewidth]{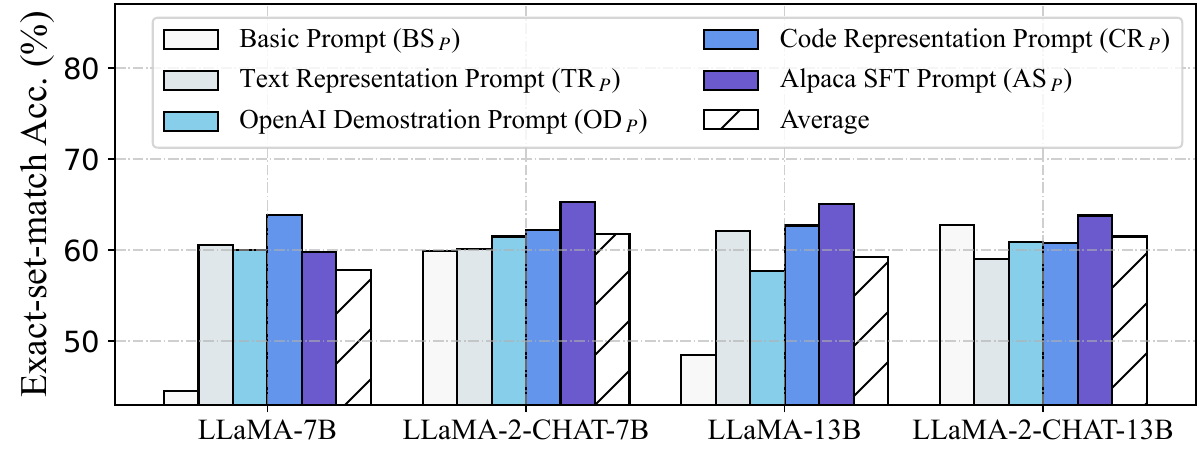}
    \includegraphics[width=0.48\linewidth]{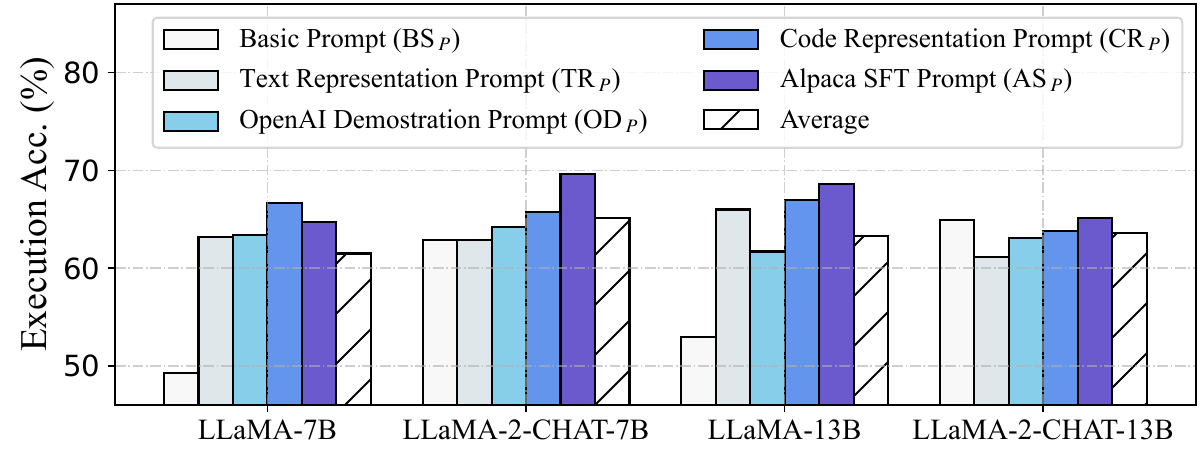}
	\caption{Zero-shot evaluation results on Spider-dev with different fine-tuned open-source LLMs.}
	\label{fig:sft_each_prompt_spider}
\end{figure*}

\subsubsection{Supervised Fine-tuning with Open-source LLM}

To further enhance Open-source LLMs' performances, we explore supervised fine-tuning for \nlsql. 
Similar to in-context learning, it may prefer different representations. 
Thus, we first fine-tune open-source LLMs on zero-shot training samples with different representations. 
Following the setting of supervised fine-tuning~\cite{instructgpt, alpaca}, we block the gradients from prompt and only update weights with those from response (SQL queries). 
We use the train split in Spider, which contains $8659$ training samples. 
For more training details, please refer to \appref{app:tuning_detail}.

\textbf{Zero-shot Scenario.}
\figref{fig:sft_each_prompt_spider} shows the performance of supervised fine-tuning with various LLMs and question representations in zero-shot scenario. 
Compared with zero-shot performance before fine-tuning in \tabref{tab:0shot_spider_ss}
, their performances are greatly enhanced. 
By comparing different representations, \alpacaprompt show obvious advantages in supervised fine-tuning as it is designed for such scenario.

We also observe the gap among different representations and model scales becomes narrow. 
The possible reason is that after fine-tuning, \revision{LLMs learn to answer new \nlsql questions without task instruction and foreign keys.} 
In this experiment, the best performance on Spider is achieved by the combination of LLaMA-13B and \alpacaprompt, whose exact-set-match and execution accuracy are $65.1\%$ and $68.6\%$. 
For more detailed numerical results, please refer to \appref{app:sft_each_prompt}. 
As for larger LLM, the combination of LLaMA-33B and \sqlprompt achieves $69.1\%$ execution accuracy and $65.9\%$ exact-set-match accuracy. 
Due to the limited resources, we leave LLMs larger than 33B as our future work. 

In summary, supervised fine-tuning is quite beneficial for open-source LLMs in \nlsql. 
Compared with OpenAI LLMs, in zero-shot scenario, fine-tuned LLaMA-13B and 30B are comparable to TEXT-DAVINCI-003 and slightly weaker than GPT-4 and GPT-3.5-TURBO. 

\begin{table}[t]
    \small
    \centering
    \tabcolsep=4.5pt
	\begin{tabular}{ll cc cc cc cc}
		\toprule
		\multirow{2}{*}{LLM}	&    \multirow{2}{*}{Org.} &	\multicolumn{2}{c}{0-shot}	&	\multicolumn{2}{c}{1-shot}	&	\multicolumn{2}{c}{3-shot}	&	\multicolumn{2}{c}{5-shot}	\\
		\cmidrule(r){3-4}
		\cmidrule(r){5-6}
		\cmidrule(r){7-8}
		\cmidrule(r){9-10}
  			&		&	EM	&	EX	&	EM	&	EX	&	EM	&	EX	&	EM	&	EX	\\
		\hline
		\multirow{3}{*}{\makecell{LLaMA\\-7B}}	&	\abfiorg	&	3.1	&	13.0	&	23.4	&	30.1	&	23.7	&	30.3	&	24.7	&	30.9	\\
            &	\absqlorg	&	3.1	&	13.0	&	13.3	&	21.4	&	15.2	&	24.1	&	15.3	&	25.0   \\
			&	\abpairorg	&	3.1	&	13.0	&	18.5	&	25.4	&	22.1	&	28.1	&	22.6	&	29.3	\\
		\hline
		\multirow{3}{*}{+ SFT}	&	\abfiorg	&	63.9	&	66.7	&	59.6	&	61.4	&	58.7	&	61.4	&	59.4	&	61.5	\\
            &  \absqlorg	&	63.9	&	66.7	&	59.8	&	62.3	&	58.8	&	61.1	&	59.5	&	62.2	\\
			&	\abpairorg	&	63.9	&	66.7	&	58.5	&	61.9	&	59.8	&	61.7	&	58.9	&	60.9	\\
		\hline
		\multirow{3}{*}{\makecell{LLaMA\\-13B}}	&	\abfiorg	&	2.4	&	20.3	&	21.6	&	33.8	&	27.3	&	38.1	&	28.5	&	38.8	\\
            &	\absqlorg	&	2.4	&	20.3	&	20.7	&	33.6	&	23.2	&	35.9	&	27.4	&	36.9  \\
			&	\abpairorg	&	2.4	&	20.3	&	13.2	&	30.0	&	15.5	&	32.3	&	16.2	&	32.4	\\
		\hline
		\multirow{3}{*}{+ SFT}	&	\abfiorg	&	62.7	&	67.0	&	61.9	&	67.1	&	60.5	&	65.0	&	60.9	&	65.0	\\
            &  \absqlorg	&	62.7	&	67.0	&	61.9	&	66.2	&	60.1	&	64.6	&	60.2	&	65.2	\\
			&	\abpairorg	&	62.7	&	67.0	&	62.5	&	66.5	&	60.6	&	66.0	&	61.3	&	66.4	\\
		\bottomrule
	\end{tabular}
  \vspace{0.1in}
	\caption{Few-shot evaluation results of supervised fine-tuned LLMs on Spider-dev.}
	\label{tab:fewshot_sft}
\end{table}

\textbf{Few-shot Scenario.}
% few-shot
After supervised fine-tuning, an important issue is: \textit{Can we continue to enhance the performance of open-source LLM by adding contextual examples?}
To answer this question, we evaluate fine-tuned LLaMA-7B and 13B with 0, 1, 3 and 5-shot prompts as shown in \tabref{tab:fewshot_sft}. 
We also add the evaluation results of original LLaMA-7B and 13B for clear comparison.
Unexpectedly, the fine-tuned LLMs fail to learn from examples. 
Specifically, adding contextual examples in test prompts incurs sudden decrease in both exact-set-match and execution match accuracy, and adding more examples is also unhelpful. 
A possible reason is that LLM overfits to zero-shot prompt, which makes examples unuseful.

In summary, open-source LLMs demonstrate significant potential for \nlsql tasks, particularly in supervised fine-tuning. 
Specifically, after fine-tuning, their performances are comparable to TEXT-DAVINCI-003 in zero-shot scenario. 
However, unlike OpenAI LLMs, fine-tuned LLMs fail to learn from contextual examples. 
The question of preserving in-context learning ability after fine-tuning remains to be explored in future studies.
\subsection{Token Efficiency}
\label{subsec:summary}

\begin{figure*}[t]
    \centering
    \subfigure[\gptfour]{\includegraphics[width=0.49\linewidth]{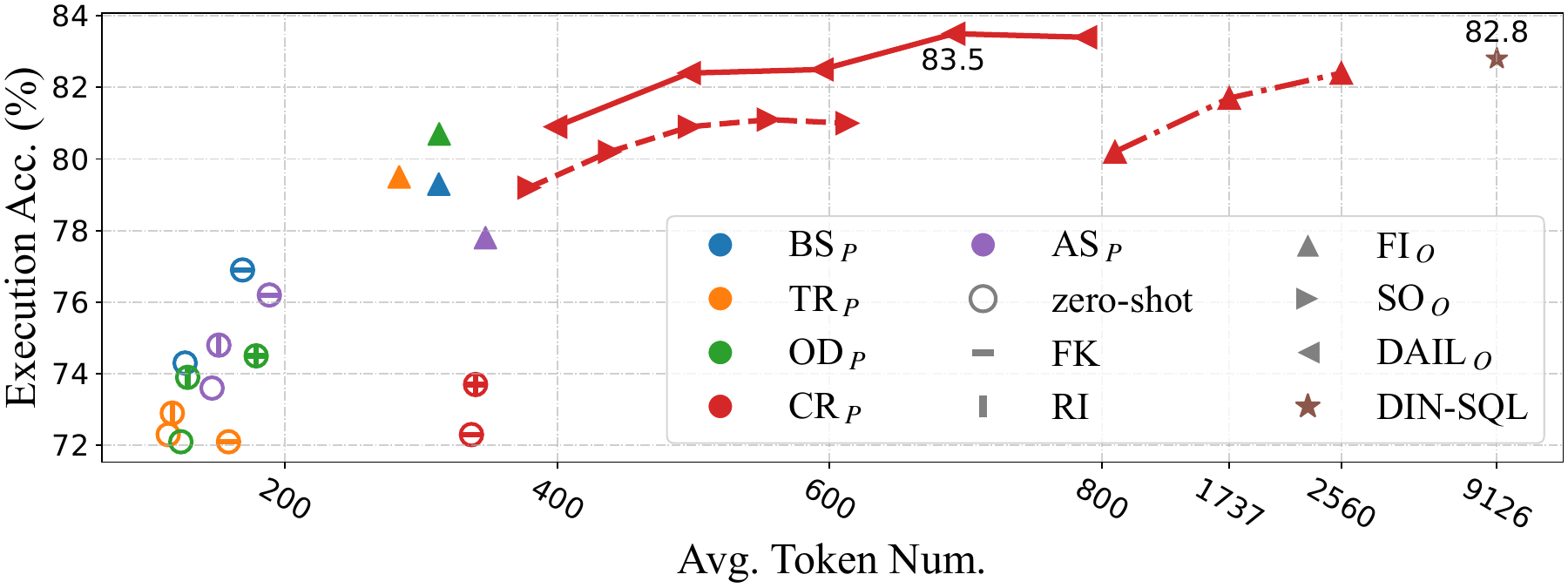}\label{fg:token_efficiency:icl:gpt4}}%
    % \hfil%
    \subfigure[\chatgpt]{\includegraphics[width=0.49\linewidth]{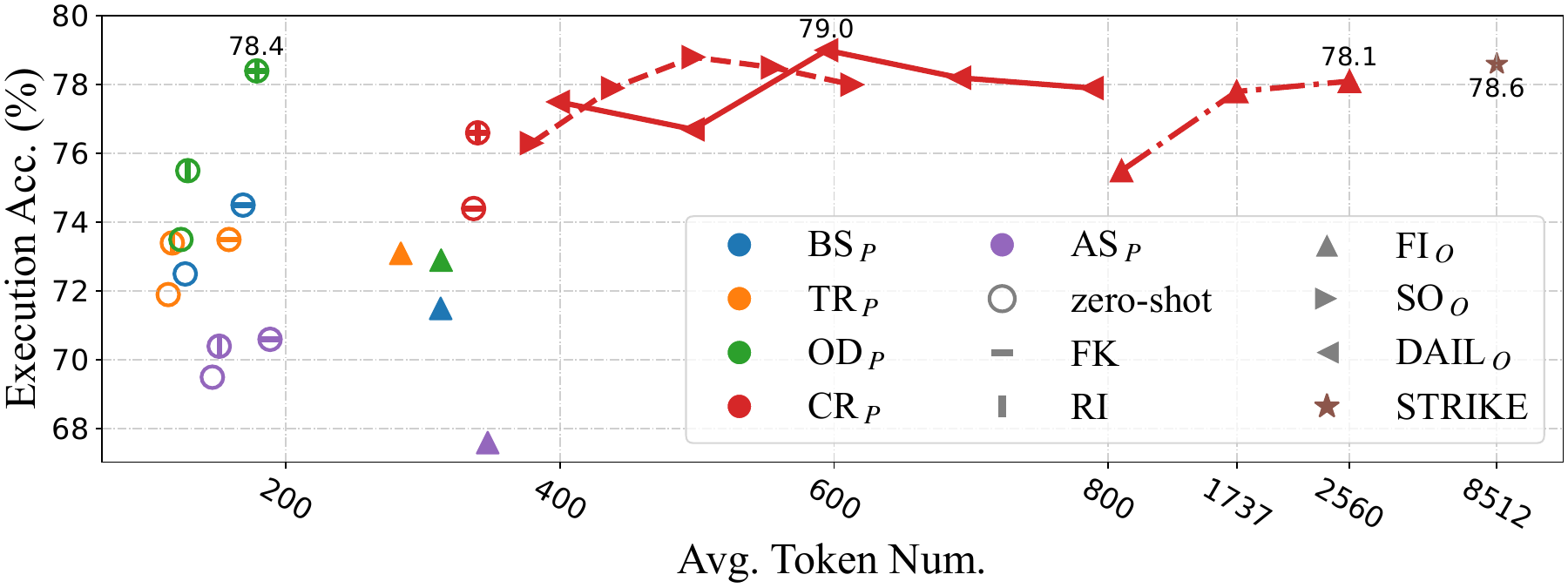}\label{fg:token_efficiency:icl:chatgpt}}%
    % \hfil%
    \hfil
    \subfigure[\davinci]{\includegraphics[width=0.49\linewidth]{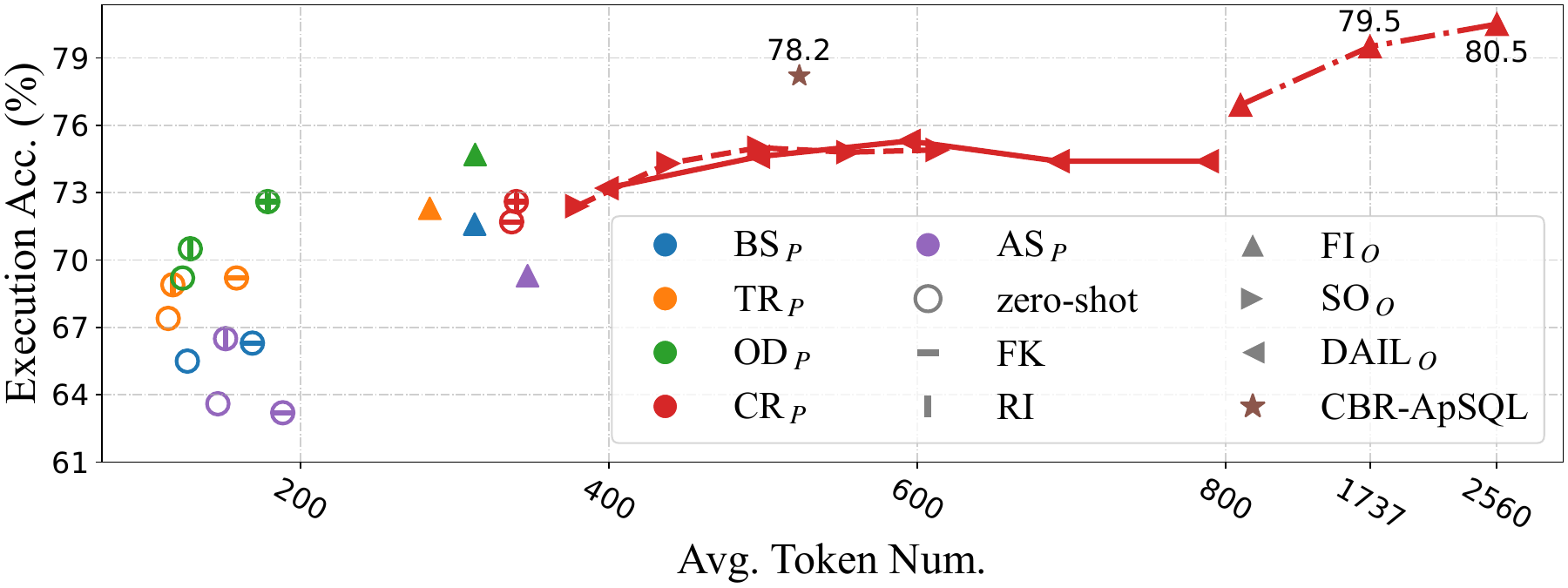}\label{fg:token_efficiency:icl:davinci}}%
    % \hfil%
    \subfigure[Open-source LLM]{\includegraphics[width=0.49\linewidth]{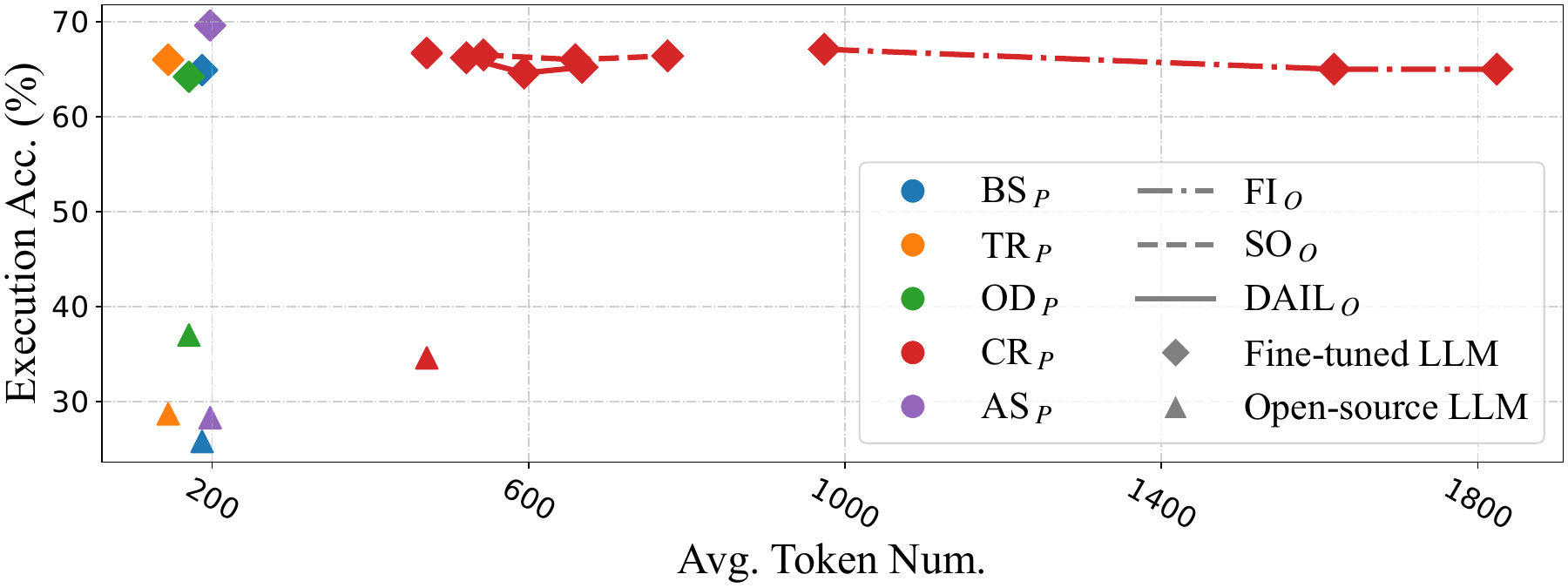}\label{fig:token_efficiency:icl:open}}
    \caption{Token efficiency of different representations in Spider-dev for OpenAI LLMs. We utilize different colors to represent different question representations and different shapes to denote different example organizations as well as the usage of foreign key information and rule implication. In particular, the overlap of shapes is used to indicate the usage of both foreign key information and rule implication. The rings stand for the prompts in zero-shot scenario and the stars stand for the previous SOTA results of few-shot methods in LLMs.}
    \label{fig:token_efficiency:icl}
\end{figure*}

Considering OpenAI LLMs are charged by token numbers, and LLMs' running time are proportional to token lengths, we underscore token efficiency in prompt engineering, which aims to achieve higher accuracy with less tokens. 
In this section, we review our experiments on Spider-dev in terms of token efficiency.  \revision{(For more efficiency analysis, please refer to \appref{eff:financial} and \ref{eff:time}.)} 
Specifically, for both OpenAI and open-source LLMs, we experimentally study the trade-off between execution accuracy and token numbers, and the token number is mainly affected by question representation and example organization. 
For example selection, we fix it as \abpqsselector.
Besides, we also include several state-of-the-art \nlsql methods in our comparison, including DIN-SQL~\cite{din-sql}, STRIKE~\cite{enhancing} and CBR-ApSQL~\cite{DBLP:journals/corr/abs-2304-13301}. 
We take their reported highest execution accuracy as their performances. 
For token cost, we average the token number of 10 randomly sampled instances for DIN-SQL. 
For STRIKE, the optimal performance are achieved by majority voting from $1$-shot to $5$-shot results, resulting in a significant increase in token cost.
Further, for CBR-ApSQL the token cost is calculated with their question representation and $8$-shot examples in \sqlorg. 

\figref{fig:token_efficiency:icl} shows the comparison in terms of token efficiency.
In zero-shot scenario, compared with rule implication, prompt with foreign keys generally achieve higher execution accuracy at the expense of more tokens. 
In few-shot scenario, comparing different organization strategies, \abfiorg are very inefficient, whose tokens numbers are several times that of \abpairorg and \absqlorg. 
Comparing \abpairorg and \absqlorg, \abpairorg together with \gptfour achieve the highest accuracy of $83.5\%$, yet having similar token cost with \absqlorg. 
Therefore, \abpairorg are more efficient than \absqlorg and \abfiorg in terms of token.

Compared with other state-of-the-art solutions, \ours outperforms DIN-SQL and STRIKE in terms of both accuracy and efficiency. 
While for CBR-ApSQL, it achieves $78.2\%$ accuracy with \davinci, but still lower than the optimal performance achieved by \abpqsselector + \abfiorg. 

Besides, For open-source LLM in \figref{fig:token_efficiency:icl:open}, the LLMs fine-tuned on \nlsql are much more efficient. 
However, as discussed in \secref{sec:sft}, adding examples is unhelpful for open-source LLMs, and even reduces their token efficiency.

In summary, token efficiency is a critical metric for real-world applications of LLMs on \nlsql. In light of this, our approach, \ours, offers a compelling solution that combines high execution accuracy with improved token efficiency. This makes it highly practical and suitable for real-world applications.

\section{Discussion}
\label{sec:discussion}
Based on our experiments, we can have some empirical insights and guidelines as follows:
\begin{itemize}
    \item For question representation, \sqlprompt and \openaiprompt are recommended, and other information such as foreign key and rule implication can be very helpful. 
    
    \item For example selection, the similarities of both natural language question and SQL query are important. 
    These two similarities together are a good indicator for designing effective selection strategy.
    
    \item For example organization, if the adopted LLM is powerful enough, like \gptfour, presenting them question and SQL query pairs is an effective yet efficient choice. 
    Otherwise, presenting them full information examples is suggested. 
    
    % \item For open-source LLM, supervised fine-tuning seems to be necessary, and \alpacaprompt used in Alpaca is suggested. 
    % Besides, to obtain comparable performance with OpenAI LLMs, larger LLMs might be helpful. 
    \item \revision{For open-source LLM, having more parameters in LLMs benefits to \nlsql task, but the training corpus plays a more crucial role. Besides, supervised fine-tuning is necessary and has considerable potential in \nlsql task.}
\end{itemize}

There are also some limitations in this paper. 
Due to limited resources, we only test two rule implications, and the exploration of more rules can further benefit LLM-based \nlsql solutions. 
\revision{We fine-tune open-source LLMs with only the Spider training set, and additional \nlsql data would further enhance LLMs.}
Besides, the databases in Spider and Spider-Realistic may be not large enough, and we believe some new challenges in effectiveness and efficiency will emerge if there are a mass of tables in \nlsql task.  
\revision{Furthermore, the current evaluation metric prioritizes correctness over efficiency, and promoting LLM to generate efficient SQL among correct alternatives remains an important, unexplored question.}
We will keep working on these limitations and open questions.

\section{Conclusions}
\label{sec:conclusion}

In this paper, we conduct a systematical study on LLM-based \nlsql from aspects of 
\revision{prompt engineering} and supervised fine-tuning. 
We point out that existing in-context learning techniques for \nlsql neglect the mapping between questions and queries, as well as the trade-off between example quality and quantity. 
To address these issues, we proposed a new prompt engineering method, named \ours, which refreshes the Spider leaderboard with $86.6\%$ execution accuracy and ranks the first place. 
Regarding supervised fine-tuning, we demonstrate the great potentials of open-source LLMs for \nlsql, underline the importance of 
\revision{training corpus} and model scaling, and point out the degeneracy of in-context learning capability after fine-tuning. 
Further, we conduct an observation over existing solutions in terms of \revision{efficiency}, which indicates \ours is much more efficient and emphasizes the importance of token efficiency in prompt engineering. 
All of these are open challenges and opportunities for future study.
We hope that our work can provide a comprehensive study about \nlsql, give some guidelines for real-world applications, and help people advance its frontiers. 

%\clearpage

\bibliographystyle{ACM-Reference-Format}
\bibliography{cite}

\clearpage

\appendix
\onecolumn
\section{Details of DAIL-SQL}

\subsection{\revision{Pseudocode}}
\label{dail-sql:pseudocode}

The pseudocode of DAIL-SQL is shown in \algref{alg:dail_sql}. In DAIL-SQL, we initially eliminate the tokens associated with databases in the questions and queries of both the target and cross-domain candidates (lines 1-10). Next, we order the candidates based on question similarity and give preference to those with a high query similarity (lines 11-20). Lastly, we choose the top $k$ candidates as examples, represent their questions and queries in the Code Representation Prompt, and feed the prompt into a large language model to generate the final predicted SQL query (lines 21-23).
 
DAIL-SQL utilizes some functions from existing works. Specifically, we identify database-related tokens in the questions using schema-linking~\cite{rat-sql} (line 2 and 4),  and extract query skeletons by retaining their SQL keywords~\cite{li2023resdsql} (line 8 and 10). In line 12, we encode the masked questions using all-mpnet-base-v2~\cite{song2020mpnet}.

\begin{algorithm}[htb]
    \caption{DAIL-SQL Algorithm}
    \label{alg:dail_sql}
    \KwIn{Target question $q$ and database $\mathcal{D}$, a set of triples $\mathcal{Q} = \{(q_i, s_i, \mathcal{D}_i)\}$, number of examples $k$, large lanuage model $\mathcal{M}$, the Code Representation $\sigma_{CR_P}$ with DAIL Organization, preliminary predictor $\mathcal{P}$, sentence embedding model $e$, and query similarity threshold $\eta$. }
    \KwOut{SQL query $s$ of the target question $q$}
    
    \textcolor{gray}{$\#\ mask\ the\ tokens\ related\ to\ databases\ in\ both\ target\ question\ and\ candidate\ questions$}
    
    $q'$ = mask\_question($q$)
    
    \For{$(q_i, s_i, D_i)\in\mathcal{Q}$}{
        $q'_i$ = mask\_question($q_i$)
    }
    
    \textcolor{gray}{$\#\ predict\ SQL\ query\ via\ preliminary\ predictor\ \mathcal{P}$}
    
    $s_\mathcal{P} = \mathcal{P}(\sigma_{CR_P}(q, \mathcal{D}))$
    
    \textcolor{gray}{$\#\ extract\ skeletons\ of\ the\ predicted\ SQL\ query\ and\ queries\ in\ candidates$}
    
    $s'_\mathcal{P}$ = extract\_skeleton($s_\mathcal{P}$)	
    
    \For{$(q_i, s_i, D_i)\in\mathcal{Q}$}{
        $s'_i$ = extract\_skeleton($s_i$)
    }
    
    \textcolor{gray}{$\#\ sort\ the\ candidates\ with\ the\ masked\ question\ similarity$}
    
    sort $\mathcal{Q}$ by cosine\_simlarity($e(q')$, $e(q'_i)$)	
    
    \textcolor{gray}{$\#\ reorder\ \mathcal{Q}\ by\ prioritizing\ the\ candidates\ with\ high\ query\ similarity$}
    
    $\mathcal{Q}_{high\_priority}, \mathcal{Q}_{low\_priority} = \emptyset, \emptyset$
    
    \For{$(q_i, s_i, D_i)\in\mathcal{Q}$} {
        \If{\rm{Jaccard\_similarity}$(s'_\mathcal{P}$, $s'_i$)$\geq\eta$}{
            $\mathcal{Q}_{high\_priority}\leftarrow(q_i, s_i, D_i)$
        }
        \Else{
            $\mathcal{Q}_{low\_priority}\leftarrow(q_i, s_i, D_i)$
        }
    }
    
    $\mathcal{Q} = \mathcal{Q}_{high\_priority} + \mathcal{Q}_{low\_priority}$
    
    \textcolor{gray}{$\#\ generate\ prompt\ and\ final\ SQL\ query$}
    
    $\mathcal{Q}_s = \mathcal{Q}[0:k]$
    
    $s_{\mathcal{M}}$ = $\mathcal{M}(\sigma_{CR_P}(q, \mathcal{D}, \mathcal{Q}_s))$ 
    
    \KwRet{	$s_{\mathcal{M}}$}
    
\end{algorithm}

\subsection{Implementation Details}
\label{app:engineering}

For question similarity in \abqsselector, \abslmselector and \abpqsselector, we first connect question words to the database with a $n$-gram matching based schema-linking method~\cite{rat-sql}. 
Then to obtain the skeleton, we replace table and column names with ``<mask>'', and values with ``<unk>''. 
At last, we embed the masked questions with a pre-trained sentence Transformer, all-mpnet-base-v2~\cite{song2020mpnet}, to calculate their similarities. 

For query similarity in \abpqsselector, we utilize Graphix~\cite{li2023graphix} as the preliminary model to generate the predicted query $s'$.
Then we obtain its skeleton by removing its database-specific information~\cite{li2023resdsql}, including column names and values. 
Finally, we calculate the Jaccard similarity between example candidate and the predicted query $s'$ as their query similarity. 
For the query similarity threshold $\tau$ in \abpqsselector, we set it as $0.9$ in the experiments of this paper.

For the submission to the Spider leaderboard, we set $\tau$ to be $0.85$ and utilize GPT-4, with \absqlprompt, \abslmselector and \abpairorg, as the preliminary model to ensure only one model involved. 
Furthermore, we process the self-consistency voting on $5$ produced queries for each question and set the argument temperature as 1.0 for variety in voting.

\section{Question Representations}

\subsection{Detailed Performance of Different Question Representations}
\label{app:prompt_spider}

The numerical results of different question representations in zero-shot scenario are show in \tabref{tab:prompt_spider}.

\begin{table*}[htb]
	\centering
	\begin{tabular}{lcccccc}
		\toprule
		\multirow{2}{*}{Prompt}	&	\multicolumn{2}{c}{\gptfour}	&	\multicolumn{2}{c}{\chatgpt}	&	\multicolumn{2}{c}{\davinci}	\\ 
        \cmidrule(r){2-3}   \cmidrule(r){4-5}  \cmidrule(r){6-7}
		&	EM	&	EX	&	EM	&	EX	&	EM	&	EX	\\	\hline
		\bsprompt	& 48.5	& {\bf 74.3}	& 45.5	& 72.5	& 29.9	& 65.5	\\
		\textprompt	& 41.4	& 72.3	& 43.2	& 71.9	& 33.7	& 67.4	\\
            \openaiprompt	& 47.5	& 73.9 	& 48.8 	& {\bf 75.5}	& 35.5	& 70.5	\\
		\sqlprompt	& 22.1	& 72.3	& 34.6	& 74.4	& 31.7	& {\bf 71.7}	\\
		\alpacaprompt	& 39.4	& 73.6 	& 37.9	& 69.5	& 23.2	& 63.6	\\
		\bottomrule
	\end{tabular}
	\caption{Details of zero-shot evaluation on Spider-dev with different question representations.}
	\label{tab:prompt_spider}
\end{table*}

\subsection{Detailed Performance of Different Representations with Foreign Keys}
\label{app:foreign_key}
The numerical results of the ablation study about foreign key information are shown in \tabref{tab:foreign_key}.

\begin{table*}[htb]
	\centering
	\begin{tabular}{lcccccc}
		\toprule
		\multirow{2}{*}{Prompt}	&	\multicolumn{2}{c}{\gptfour}	&	\multicolumn{2}{c}{\chatgpt}	&	\multicolumn{2}{c}{\davinci}	\\
        \cmidrule(r){2-3}   \cmidrule(r){4-5}  \cmidrule(r){6-7}
		&	EM	&	EX	&	EM	&	EX	&	EM	&	EX	\\	\hline
            \bsprompt\wfk	& 48.4(-0.1)	& 76.9(+2.6)	& 47.9(+2.4)	& 74.5(+2.0)	& 32.4(+2.5)	& 66.3(+0.8)	\\
            \textprompt\wfk	& 41.9(+0.5)	&  72.1(-0.2)	& 44.3(+1.1)	& 73.5(+1.6)	& 34.1(+0.4)	& 69.2(+1.8)	\\
		\openaiprompt\wfk	& 46.1(-1.4)	& 74.5(+0.6) 	& 51.5(+2.7) 	& 78.4(+2.9)	& 34.3(-1.2)	& 72.6(+2.1)	\\
		\alpacaprompt\wfk	& 39.7(+0.3)	& 76.2(+2.6) 	& 38.9(+1.0)	& 70.6(+1.1) 	& 23.0(-0.2)	&  63.2(-0.4)	\\
		\bottomrule
	\end{tabular}
	\caption{Details of zero-shot evaluation on Spider-dev with foreign keys and comparisons with the results obtained without foreign keys in Table~\ref{tab:prompt_spider}.}
	\label{tab:foreign_key}
\end{table*}

\subsection{Detailed Performance of Different Representations with/without Explanation rule}
\label{app:rule}

The numerical results of ablation study about rule implication are shown in \tabref{tab:rule}.

\begin{table*}[htb]
	\centering
	\begin{tabular}{lcccccc}
		\toprule
		\multirow{2}{*}{Prompt}	&	\multicolumn{2}{c}{\gptfour}	&	\multicolumn{2}{c}{\chatgpt}	&	\multicolumn{2}{c}{\davinci}	\\
        \cmidrule(r){2-3}   \cmidrule(r){4-5}  \cmidrule(r){6-7}
		&	EM	&	EX	&	EM	&	EX	&	EM	&	EX	\\	\hline
		\textprompt\wrule	& 43.5(+2.1)	& 72.9(+0.6)	& 46.4(+3.2)	& 73.4(+1.5)	& 36.8(+3.1)	& 68.9(+1.5)	\\
            \openaiprompt\worule	& 42.7(-4.8)	& 72.1(-1.8) 	& {\bf 42.5(-6.3)} 	& 73.5(-2.0)	& 33.1(-2.4)	& 69.2(-1.3)	\\
		\sqlprompt\wrule	& 26.6(+4.5)	& 73.7(+1.4)	& 37.9(+3.3)	& 76.6(+2.2)	& 36.8(+5.1)	& 72.6(+0.9)	\\
		\alpacaprompt\wrule	& 43.0(+3.6)	& 74.8(+1.2)	& 40.2(+2.3)	& 70.4(+0.9)	& 27.1(+3.9)	& {\bf 66.5(+2.9)}	\\
		\bottomrule
	\end{tabular}
	\caption{Details of zero-shot evaluation on Spider-dev with/without rule implication ``with no explanation'' in instructions and comparisons with their opposites in Table~\ref{tab:prompt_spider}.}
	\label{tab:rule}
\end{table*}

\subsection{Question Representations with Rule Implication ``Let's think step by step''} 
\label{app:step_by_step}

The numerical results of the ablation study with the rule ``Let's think step by step'' are shown in \tabref{tab:step_by_step}.

\begin{table}[htb]
	\centering
	\begin{tabular}{ccccc}
		\toprule
		\multirow{2}{*}{Prompt}	&	\multicolumn{2}{c}{\chatgpt}	&	\multicolumn{2}{c}{\davinci}	\\
        \cmidrule(r){2-3}   \cmidrule(r){4-5}   
		&	EM	&	EX	&	EM	&	EX	\\	\hline
		\textprompt\wrule	& 20.0(-23.2)	& 45.9(-26.0)	& 23.0(-10.7)	& 46.1(-21.3)	\\
		\sqlprompt\wrule	& 21.7(-12.9)	& 52.2(-22.2)	& 31.9(+0.2)	& 63.8(-7.9)	\\
		\openaiprompt\wrule	& 48.4(-0.4) 	& 75.8(+0.3)	& 41.2(+5.7)	& 72.4(+1.9)	\\
		\alpacaprompt\wrule	 & 21.5(-16.4)	& 49.3(-20.2)	& 27.9(+4.7)	& 64.4(+0.8)	\\
		\bottomrule
	\end{tabular}
	\caption{Zero-shot evaluation results on Spider-dev with "Let's think step by step" rule implication in instructions and comparisons with Table~\ref{tab:prompt_spider}.}
	\label{tab:step_by_step}
\end{table}

\section{In-context Learning for \nlsql}

\begin{figure*}[htb]
    \centering
    \includegraphics[width=0.48\linewidth]{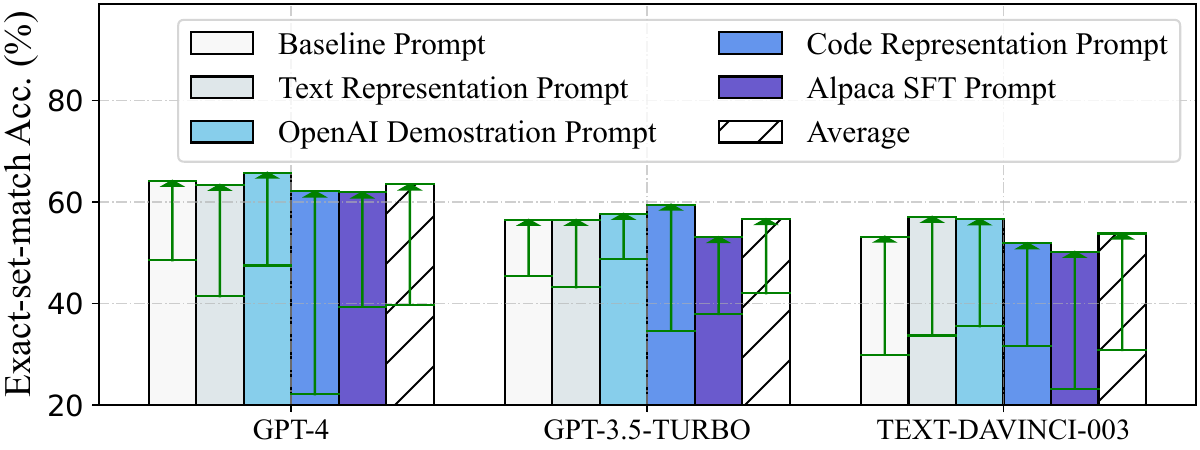}
    \includegraphics[width=0.48\linewidth]{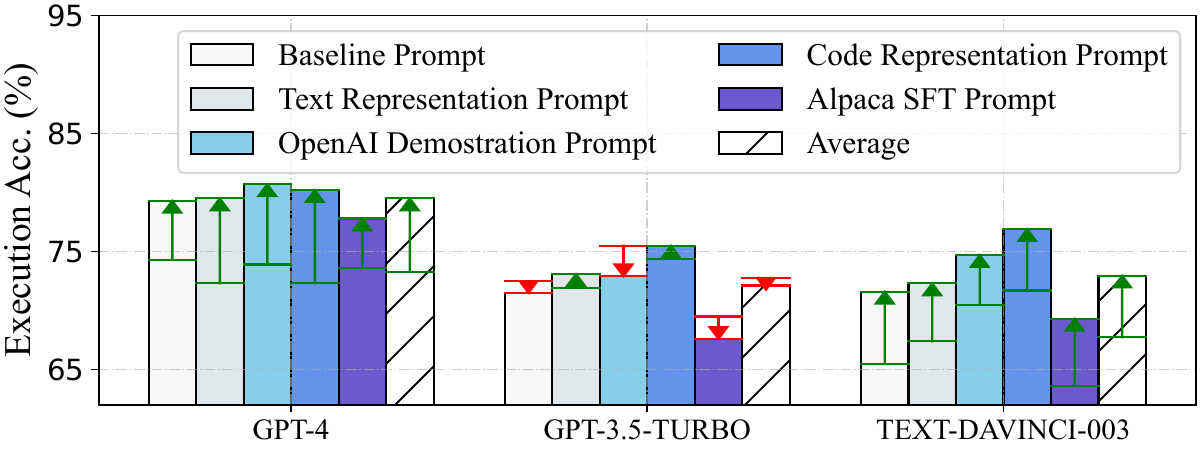}
    \caption{Results of one-shot evaluation on Spider-dev with different question representations and comparisons with the results of zero-shot evaluation in \figref{fig:openai_zero_shot}. The \textcolor{green}{green} arrow indicates increase, and \textcolor{red}{red} arrow indicates decrease.}
	\label{fig:prompt_spider_1shot}
\end{figure*}

In~\figref{fig:prompt_spider_1shot}, we present the results of our one-shot evaluation for different question representations. 
Specifically, we use \abfiorg examples with \abpqsselector here, and \tabref{tab:prompt_spider_1shot} shows the comparisons with the zero-shot scenario. 
By comparing zero-shot and one-shot evaluation results, adding contextual example show obvious and consistent improvements for all LLMs in exact-set-match accuracy. 
In term of execution accuracy, contextual examples benefits both \gptfour and \davinci. 
However, for \chatgpt, adding contextual examples only benefits \abtextprompt and \absqlprompt, indicating the in-context learning capability bias in different LLMs. 
By comparing different representations, \absqlprompt shows obvious advantage in execution accuracy, as taking the advantage of programming. 

\subsection{One-Shot Evaluation on Different Question Representation}
\label{app:prompt_spider_1shot}

\begin{table*}[htb]
	\centering
	\begin{tabular}{lcccccc}
		\toprule
		\multirow{2}{*}{Prompt}	&	\multicolumn{2}{c}{\gptfour}	&	\multicolumn{2}{c}{\chatgpt}	&	\multicolumn{2}{c}{\davinci}	\\
        \cmidrule(r){2-3}   \cmidrule(r){4-5}   \cmidrule(r){6-7}
		&	EM	&	EX	&	EM	&	EX	&	EM	&	EX	\\	\hline
		\bsprompt	& 64.2(+15.7)	& 79.3(+5.0)	& 56.5(+11.0)	& 71.5(-1.0)	& 53.2(+23.3)	& 71.6(+6.1)	\\
		\textprompt	& 63.4(+22.0) 	& 79.5(+7.2)	& 56.5(+13.3)	& 73.1(+1.2)	& 57.1(+23.4)	& 72.3(+4.9) 	\\
            \openaiprompt	& 65.8(+18.3)	& {\bf 80.7(+6.8)}	& 57.7(+8.9)	& 72.9(-2.6)	& 56.6(+21.1)	& 74.7(+4.2)	\\
		\sqlprompt	& 62.1(+40.0)	& 80.2(+7.9)	& 59.5(+24.9)	& {\bf 75.5(+1.1)}	& 51.9(+20.2)	& {\bf 76.9(+5.2)}	\\
		\alpacaprompt	& 61.9(+22.5)	& 77.8(+4.2)	& 53.1(+15.2)	& 67.6(-1.9)	& 50.2(+27.0)	& 69.3(+5.7)	\\
		\bottomrule
	\end{tabular}
	\caption{Details of one-shot evaluation on Spider-dev with different question representations and comparisons with the results of zero-shot evaluation in Table~\ref{tab:prompt_spider}.}
	\label{tab:prompt_spider_1shot}
\end{table*}

\subsection{Detailed Performance of Different Example Organizations}
\label{app:exp_organization}

The numerical results of different example organization strategies in few-shot scenario are shown in~\tabref{tab:k_prompt_spider} and~\ref{tab:k_prompt_spider_realistic}.

\begin{table*}[htb]
	\centering
	\begin{tabular}{cc|cccccc}
		\toprule
		\multirow{2}{*}{Few-shot}	& \multirow{2}{*}{Presentation} &	\multicolumn{2}{c}{\gptfour}	&	\multicolumn{2}{c}{\chatgpt}	&	\multicolumn{2}{c}{\davinci}	\\
		
		& & EM	&	EX	& EM	&	EX	&	EM	&	EX	\\	\hline
            0-shot  & -   &  22.1   & 72.3  & 34.6  & 74.4  & 31.7  & 71.7  \\  \hline
		\multirow{3}{*}{1-shot}	&	\fiorg	 &  62.1  & 80.2  & 59.5  & 75.5  & 51.9	& 76.9				\\
		&	\sqlorg	& 55.2  & 79.2  & 51.2 & 76.3  & 41.2	& 72.4		\\
		&	\pairorg	& 62.9  & 80.9  & 57.6  & 77.5  & 46.9	& 73.2				\\	\hline
		\multirow{3}{*}{3-shot}	&	\fiorg	& 69.1  & 81.7  & 63.9  & 77.8	& 64.4  & 79.5				\\
		&	\sqlorg		& 64.7  & 80.2  & 56.2  & 77.9  & 48.6	& 74.3		\\
		&	\pairorg	& 69.0  & 82.4  & 61.9  & 76.7 & 54.0	& 74.6		\\	\hline
		\multirow{3}{*}{5-shot}	&	\fiorg	& 71.9  & 82.4  & 66.7   & 78.1	& 67.7  & 80.5				\\
		&	\sqlorg		&  66.6 & 80.9  & 56.2  & 78.8  & 52.1	& 75.0				\\
		&	\pairorg	& 70.8  & 82.5  & 64.3  & 79.0	& 58.2   & 75.3				\\	\hline
		\multirow{2}{*}{7-shot}	&	\sqlorg		& 67.8  & 81.1  & 57.2    & 78.5   & 52.1	& 74.8			\\
		&	\pairorg	& 72.5  & 83.5  & 65.6   & 78.2	& 59.0   & 74.4			\\ \hline
		\multirow{2}{*}{9-shot}	&	\sqlorg		& 67.6  & 81.0  & 57.7  & 78.0  & 53.0	& 74.9				\\
		&	\pairorg	& 72.8  & 83.4  & 65.3  & 77.9	& 60.3   & 74.4				\\
		\bottomrule
	\end{tabular}
	\caption{Details of Few-shot evaluation on Spider development split with different organizations.}
	\label{tab:k_prompt_spider}
\end{table*}

\begin{table*}[htb]
	\centering
	\begin{tabular}{cc|cccccc}
		\toprule
		\multirow{2}{*}{Few-shot}	& \multirow{2}{*}{Presentation} &	\multicolumn{2}{c}{\gptfour}	&	\multicolumn{2}{c}{\chatgpt}	&	\multicolumn{2}{c}{\davinci}	\\
		
		& & EM	&	EX	& EM	&	EX	&	EM	&	EX	\\	\hline
            0-shot  & -   &  19.9   & 66.5  & 29.3  & 67.3  & 28.0  & 65.0  \\  \hline
		\multirow{3}{*}{1-shot}	&	\fiorg	 &  50.2  & 73.2  & 48.6  & 69.1  & 42.1	& 67.1				\\
		&	\sqlorg	& 45.3  & 70.7  & 42.7  &  68.7  & 34.6	& 64.4	\\
		&	\pairorg	& 51.2  & 72.2  & 48.4  & 69.3  & 38.6	& 64.6			\\	\hline
		\multirow{3}{*}{3-shot}	&	\fiorg	& 59.1  & 73.6  & 52.6  & 68.3	& 54.5  & 69.7			\\
		&	\sqlorg		& 56.9  & 73.8  & 48.8  & 70.9  & 42.7	& 65.9	\\
		&	\pairorg	& 60.8  & 74.8  & 52.8  & 69.1 & 48.4	& 65.7		\\	\hline
		\multirow{3}{*}{5-shot}	&	\fiorg	& 63.8  & 75.4  & 58.7   & 70.1	& 57.7  & 70.1			\\
		&	\sqlorg		&  58.9 & 74.4  & 49.2  & 71.3  & 45.3	& 66.1			\\
		&	\pairorg	& 60.0  & 75.0  & 55.3  & 69.3	& 51.4   & 65.2			\\	\hline
		\multirow{2}{*}{7-shot}	&	\sqlorg		& 59.6  & 74.6  & 50.0    & 70.1   & 49.0	& 69.3			\\
		&	\pairorg	& 63.6  & 75.8  & 57.1  & 67.9	& 51.6   & 64.0 	\\ \hline
		\multirow{2}{*}{9-shot}	&	\sqlorg		& 60.0  & 75.6  & 49.8  & 70.9  & 48.8	& 68.7				\\
		&	\pairorg	& 63.8  & 76.0  & 57.5  & 67.9	& 53.7   & 65.2			\\
		\bottomrule
	\end{tabular}
	\caption{Details of Few-shot evaluation on Spider-Realistic dataset with different organizations.}
	\label{tab:k_prompt_spider_realistic}
\end{table*}

\subsection{\revision{Comparisons between DAIL-SQL and Previous Works}}
\label{app:exp_cmp_plm_rule}

In this sub section, we compare DAIL-SQL with previous works, including works based on parsing rules, pre-trained language models (PLM), and large language models (LLM). As Table~\ref{tab:dif_base} shows, the performance of PLM-based and rule-base methods has recently become less competitive. Consequently, LLM-based methods have emerged as a promising and dominant approach for \nlsql tasks. Given this shift, in this paper, we focus more on a systematically study for LLM-based \nlsql methods.

\begin{table}[htb]
	\centering
	\begin{tabular}{ll cc}
		\toprule
		Classification & Method & Dev. EX (\%) & Test EX (\%) \\
		\hline
		\multirow{2}{*}{rule-based} & Duoquest~\cite{DBLP:conf/sigmod/BaikJCJ20} & 63.5 & 63.5 \\
		& ATHENA++~\cite{DBLP:journals/pvldb/SenLQOEDSMSS20} & 78.82 & - \\
		\hline 
		\multirow{8}{*}{PLM-based} & ValueNet~\cite{valuenet} & 67.0 & - \\
		& BRIDGE v2 + BERT~\cite{DBLP:conf/emnlp/LinSX20} & 68.0 & 64.3 \\
            & T5-Base~\cite{scholak2021picard} & 57.9 & - \\
		& T5-Large~\cite{scholak2021picard} & 67.2 & - \\
		& T5-3B~\cite{scholak2021picard} & 74.4 & 70.1 \\
		& T5-3B + PICARD~\cite{scholak2021picard} & 79.3 & 75.1 \\
		& RESDSQL-3B + NatSQL~\cite{li2023graphix} & 84.1 & 79.9 \\
            & Fine-tuned BERT (ours)  &   53.6    &   -  \\
		\hline
		\multirow{4}{*}{LLM-based} & C3 + ChatGPT + Zero-Shot~\cite{dong2023c3} & 81.8 & 82.3 \\
		& DIN-SQL + GPT-4~\cite{din-sql} & 82.8 & 85.3 \\
		& DAIL-SQL + GPT-4 & 83.1 & 86.2 \\
		& DAIL-SQL + GPT-4 + Self-consistency & 83.6 & \bf{86.6} \\
		\bottomrule
	\end{tabular}
	\caption{Reported execution accuracy (EX) of various solutions on development set (Dev.) and test set (Test) of Spider, including LLM-based, PLM-based, and rule-based approaches.}
	\label{tab:dif_base}
\end{table}

\section{Supervised fine-tuning for \nlsql}
\subsection{Detailed Performance of Open-source LLMs on Spider-Realistic}
\label{app:0shot_realistic_ss}

The numerical results are shown in \tabref{tab:0shot_realistic_ss}.

\begin{table*}[htb]
	\centering
	\begin{tabular}{llcc cc cc cc cc cc}
		\toprule
		\multirow{2}{*}{Stage}	& 	\multirow{2}{*}{LLM}	&	\multicolumn{2}{c}{\abbsprompt}	&	\multicolumn{2}{c}{\abtextprompt}	&	\multicolumn{2}{c}{\abopenaiprompt}	&	\multicolumn{2}{c}{\absqlprompt}	&	\multicolumn{2}{c}{\abalpacaprompt}	&	\multicolumn{2}{c}{Average}	\\
        \cmidrule(r){3-4} \cmidrule(r){5-6} \cmidrule(r){7-8} \cmidrule(r){9-10} \cmidrule(r){11-12}    \cmidrule(r){13-14}
		&		&	EM	&	EX	&	EM	&	EX	&	EM	&	EX	&	EM	&	EX	&	EM	&	EX    &   EM   &	EX	\\	\hline
  \multirow{3}{*}{Pre-training}	&	LLaMA-7B	&	4.7	&	12.0	&	1.4	&	4.9	&	1.4	&	5.1	&	3.1	&	13.0	&	0.2	&	2.8	&	2.2	&	7.6	\\
	&	LLaMA-13B	&	5.9	&	15.7	&	3.3	&	13.6	&	4.3	&	17.9	&	2.4	&	20.3	&	3.1	&	13.8	&	3.8	&	16.3	\\
	&	LLaMA-33B	&	7.1	&	17.5	&	8.3	&	21.5	&	9.6	&	28.3	&	7.9	&	\textbf{34.6}	&	\textbf{10.8}	&	25.2	&	8.7	&	25.4	\\
	\hline																				
\multirow{7}{*}{Aligned}	&	Alpaca-7B	&	7.7	&	22.8	&	9.6	&	19.3	&	11.4	&	21.7	&	12.8	&	26.0	&	0.8	&	6.9	&	8.5	&	19.3	\\
	&	GPT4ALL-7B	&	3.5	&	12.6	&	7.7	&	19.3	&	6.1	&	18.5	&	7.5	&	17.1	&	1.4	&	6.5	&	5.2	&	14.8	\\
	&	LLaMA-2-CHAT-7B	&	11.4	&	21.7	&	5.1	&	12.0	&	7.5	&	14.4	&	7.5	&	17.7	&	3.7	&	13.6	&	7.0	&	15.9	\\
	&	LLaMA-2-CHAT-13B	&	14.4	&	25.8	&	12.8	&	26.4	&	11.6	&	22.0	&	\textbf{17.9}	&	32.9	&	15.9	&	28.3	&	14.5	&	27.1	\\
	&	Vicuna-7B	&	6.7	&	18.3	&	0.8	&	8.1	&	4.7	&	16.3	&	3.9	&	14.0	&	0.6	&	4.9	&	3.3	&	12.3	\\
	&	Vicuna-13B	&	6.9	&	19.3	&	4.9	&	16.1	&	8.5	&	25.2	&	4.3	&	27.6	&	4.1	&	15.0	&	5.7	&	20.6	\\
	&	Vicuna-33B	&	8.1	&	20.7	&	13.8	&	28.7	&	16.9	&	\textbf{37.0}	&	5.1	&	34.3	&	8.7	&	27.6	&	10.5	&	30.0	\\
		% RLHF	&	StableVicuna-13B	&	24.2	&	10.0	&	24.4	&	9.8	&	29.9	&	9.8	&	28.7	&	15.4	&	12.6	&	2.8	&	24.0	&	9.6	\\
		\bottomrule
	\end{tabular}
	\caption{Zero-shot evaluation results on Spider-Realistic with different open-source LLMs.}	
	\label{tab:0shot_realistic_ss}
\end{table*}

\subsection{Detailed Performance of Open-source LLMs on Spider-dev in Few-shot Scenario}
\label{app:kshot_llms}

The numerical results are shown in \tabref{tab:kshot_llms}.

\begin{table*}[htb]
	\begin{tabular}{c c cc cc cc cc}
		\toprule
		\multirow{2}{*}{Example Organization}  &   \multirow{2}{*}{LLM}	&	\multicolumn{2}{c}{0-shot}  &	\multicolumn{2}{c}{1-shot}	&	\multicolumn{2}{c}{3-shot}	&	\multicolumn{2}{c}{5-shot}	\\
		\cmidrule(r){3-4}
		\cmidrule(r){5-6}
		\cmidrule(r){7-8}
		\cmidrule(r){9-10}
		&  &	EM & EX	&	EM & EX   &   EM  &   EX  & EM & EX	\\
		\hline
  		\multirow{2}{*}{\sqlorg}	&	LLaMA-33B	&	12.2	&	42.8	&	24.0	&	42.5	&	28.2	&	45.7	&	31.3	&	46.8	\\
			&	Vicuna-33B	&	6.9	&	43.7	&	13.3	&	46.7	&	17.1	&	49.5	&	19.5	&	49.5	\\
		\hline																			
		\multirow{2}{*}{\pairorg}	&	LLaMA-33B	&	12.2	&	42.8	&	28.5	&	46.4	&	34.9	&	47.9	&	34.5	&	45.8	\\
			&	Vicuna-33B	&	6.9	&	43.7	&	18.7	&	45.5	&	26.3	&	49.1	&	28.6	&	50.2	\\
		\hline																			
		\multirow{2}{*}{\fiorg}	&	LLaMA-33B	&	12.2	&	42.8	&	30.1	&	46.8	&	35.1	&	48.9	&	36.4	&	50.2	\\
			&	Vicuna-33B	&	6.9	&	43.7	&	22.1	&	49.2	&	27.4	&	49.9	&	28.0	&	51.1	\\
		\bottomrule
	\end{tabular}
	\caption{Detailed performance of open-source LLMs on Spider-dev in few-shot scenario.}
	\label{tab:kshot_llms}
\end{table*}

\subsection{\revision{Experiments in Few-shot Scenario for Different Model Sizes and Training Corpus of Open-source LLMs}}
\label{app:kshot_llms_size_and_corpus}

 We also evaluate the LLaMA-2-CHAT-70B, Falcon-40B, CodeLLaMA-34B and Vicuna-33B in few-shot scenario, and summarize their performance in \tabref{tab:larger_llm}. Similar with the situation in zero-shot scenario, LLaMA-2-CHAT-70 outperforms Vicuna-33B with $59.0\%$ execution accuracy, whereas Falcon-40B reaches only $19.1\%$ that is overwhelmed by CodeLLaMA-34B. And again, CodeLLaMA-34B outperforms LLaMA-2-CHAT-70B by $12.4\%$ in the few-shot scenario. This further confirms that having more parameters is beneficial, meanwhile, these results also underscore the critical importance of the training corpus.

\begin{table}[htb]
    \centering
    \begin{tabular}{l cc}
    \toprule
    LLM & EM & EX \\
    \hline
    LLaMA-2-CHAT-70B & 42.9 & 59.0 \\
    Falcon-40B & 4.8 & 19.1 \\
    CodeLLaMA-34B & 59.9 & 71.4  \\
    Vicuna-33B & 31.3 & 51.7 \\
    \bottomrule
    \end{tabular}
    \caption{Performances of different open-source LLMs with DAIL-SQL.}
    \label{tab:larger_llm}
\end{table}

\subsection{Details for Supervised Fine-tuning}
\label{app:tuning_detail}

For dataset, we use the train split in Spider, which totally contains $8659$ training samples. 
For hyper-parameters, we set global batch size as $256$, and search learning rate in $[1e-6, 1e-4]$ and weight decay in $\{1, 0.1, 0.01, 0\}$.  
During fine-tuning, we use a cosine learning rate scheduler in transformers~\cite{hftransformers} with a warm ratio $0.03$. 
Besides, all LLMs are fine-tuned on a server with eight 64G A100 GPUs.

\subsection{Detailed Performance of Fine-tuned Open-source LLM on Spider and Spider-Realistic}
\label{app:sft_each_prompt}

The numerical results on Spider-dev and Spider-Realistic are all shown in \tabref{tab:sft_each_prompt}. 

\begin{table*}[htb]
	\centering
	\begin{tabular}{cc ccccc c ccccc c}
		\toprule
		\multirow{2}{*}{LLM}	&	\multirow{2}{*}{Metric}	&	\multicolumn{6}{c}{Spider}	&	\multicolumn{6}{c}{Spider-Realistic}\\
		\cmidrule(r){3-8} \cmidrule(r){9-14}
		
		&	&	\abbsprompt	&	\abtextprompt		&	\abopenaiprompt	&	\absqlprompt	&	\abalpacaprompt	&	Average   &   \abbsprompt	&	\abtextprompt		&	\abopenaiprompt	&	\absqlprompt	&	\abalpacaprompt	&   Average \\
		\cmidrule(r){1-8} \cmidrule(r){9-14}
  		\multirow{2}{*}{LLaMA-7B}	&	EM	&	44.5	&	60.6	&	60.0	&	\textbf{63.9}	&	59.8	&	57.8 &   33.3	&	45.5	&	43.9	&	\textbf{51.6}	&	46.5    &   44.2	\\
		&	EX	&	49.3	&	63.2	&	63.4	&	\textbf{66.7}	&	64.7	&  61.5    &	35.0	&	48.8	&	47.4	&	\textbf{53.7}	&	46.5  &   46.3	\\
	\cmidrule(r){1-8} \cmidrule(r){9-14}
    \multirow{2}{*}{LLaMA-2-CHAT-7B}    &   EM    &   59.9    &   60.1    &   61.5    &   62.2    &   \textbf{65.3}    & 61.8    &  48.2    &		45.7    &		45.3    &		49.4    &		\textbf{52.2}    &   48.2 \\
    &   EX  &   62.9    &   62.9    &   64.2    &   65.7    &   \textbf{69.6}    &   65.1    &    51.4    &		44.5    &		47.4    &		49.6    &		\textbf{54.9}    & 49.6\\
    \cmidrule(r){1-8} \cmidrule(r){9-14}
	\multirow{2}{*}{LLaMA-13B}	&	EM	&	48.5	&	62.1	&	57.7	&	62.7	&	\textbf{65.1}	&	59.2 &   39.0	&	50.4	&	48.2	&	48.2	&	\textbf{52.2}	&  47.6    \\
		&	EX	&	53.0	&	66.0	&	61.7	&	67.0	&	\textbf{68.6}	&  63.3   &	44.7	&	54.1	&	51.6	&	52.6	&	\textbf{54.7}  &   51.5	\\
    \cmidrule(r){1-8} \cmidrule(r){9-14}
    \multirow{2}{*}{LLaMA-2-CHAT-13B}   &   EM   &   62.8   &		59.0   &		60.9   &		60.8   &		\textbf{63.8}   &   61.5    &   \textbf{53.7}   &	47.0   &	47.2   & 49.6   &	\textbf{53.7}   &	50.2 \\
    &   EX   &  64.9   &    61.1   &	63.1   &	63.8   &	\textbf{65.1}  &   63.6   &   \textbf{54.5}   &	47.6   &	48.6   &	51.2   &	52.8   &	50.9 \\
		\bottomrule
	\end{tabular}
	\caption{Performance of supervised fine-tuning on Spider and Spider-Realistic with respect to different representations and LLMs.}
	\label{tab:sft_each_prompt}
\end{table*}

\section{Efficiency}

\subsection{\revision{Financial Efficiency}}
\label{eff:financial}

\begin{figure}[htb]
    \centering
    \subfigure[\gptfour]{\includegraphics[width=0.49\linewidth]{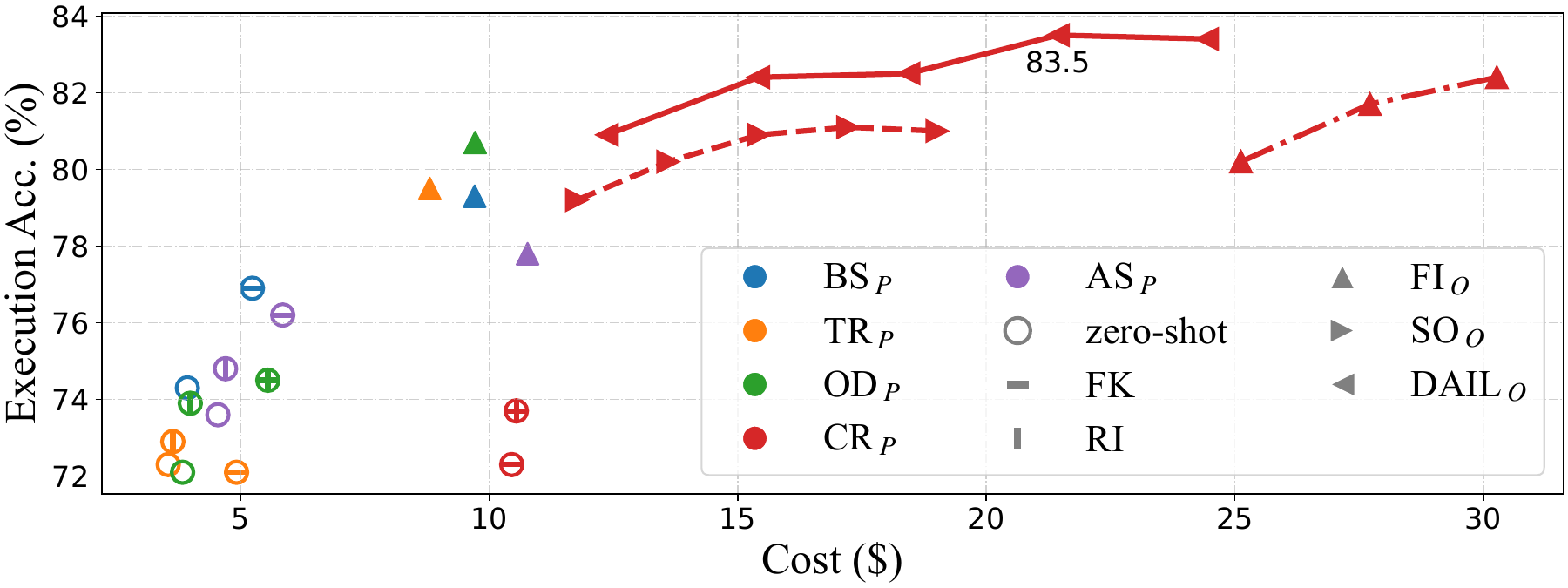}\label{fig:price:gpt4}}
    \subfigure[\chatgpt]{\includegraphics[width=0.49\linewidth]{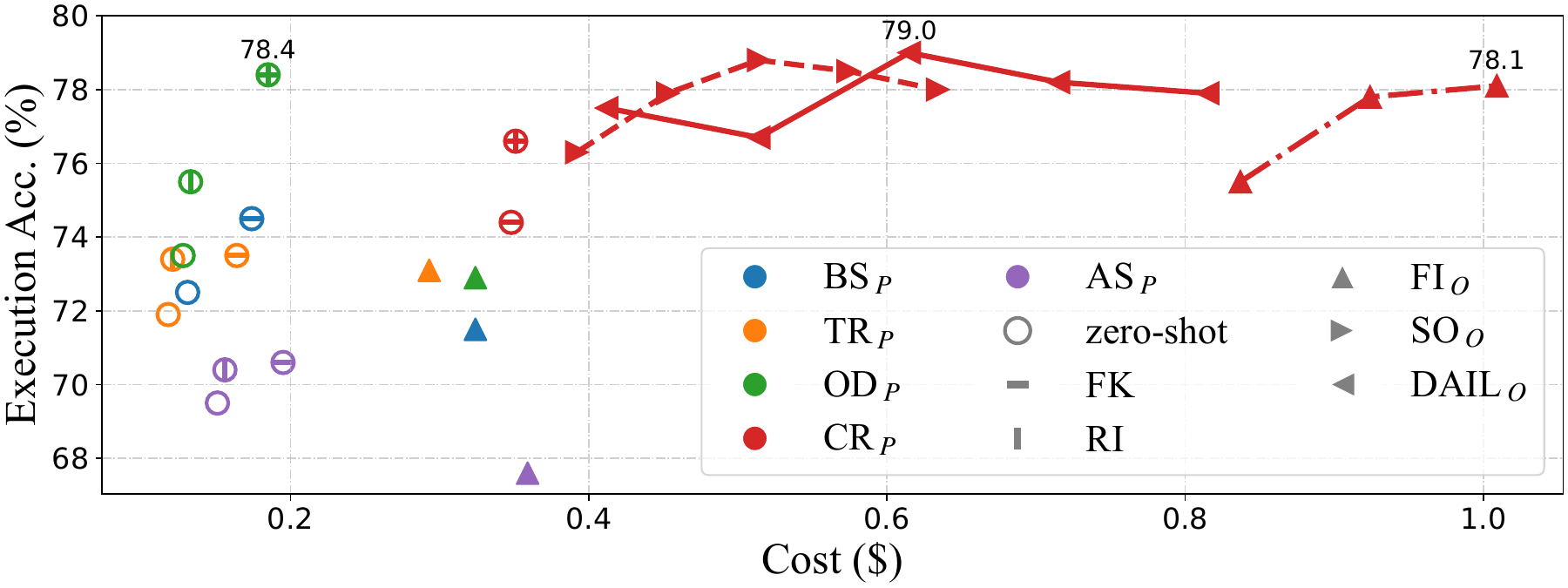}\label{fig:price:chatgpt}}
    
    \caption{Price of API calling to inference questions on Spider-dev for \gptfour and \chatgpt. We utilize different colors to represent different question representations and different shapes to denote different example organizations as well as the usage of foreign key information and rule implication. In particular, the overlap of shapes is used to indicate the usage of both foreign key information and rule implication. The rings stand for the prompts in zero-shot scenario.}
    \label{fig:price}
\end{figure}

We estimate the financial expenses for each experiment with \gptfour and \chatgpt based on the API price reported on November 8, 2023, as illustrated in Figure~\ref{fig:price}. The total costs amount to 323.2\$ and 12.0\$ for \gptfour and \chatgpt, respectively.

\subsection{\revision{Time-Consuming}}
\label{eff:time}

\begin{figure}[htb]
    \centering

    \subfigure[Averaged token number of each prompt and their corresponding performance on Spider-dev.]{\includegraphics[width=0.49\linewidth]{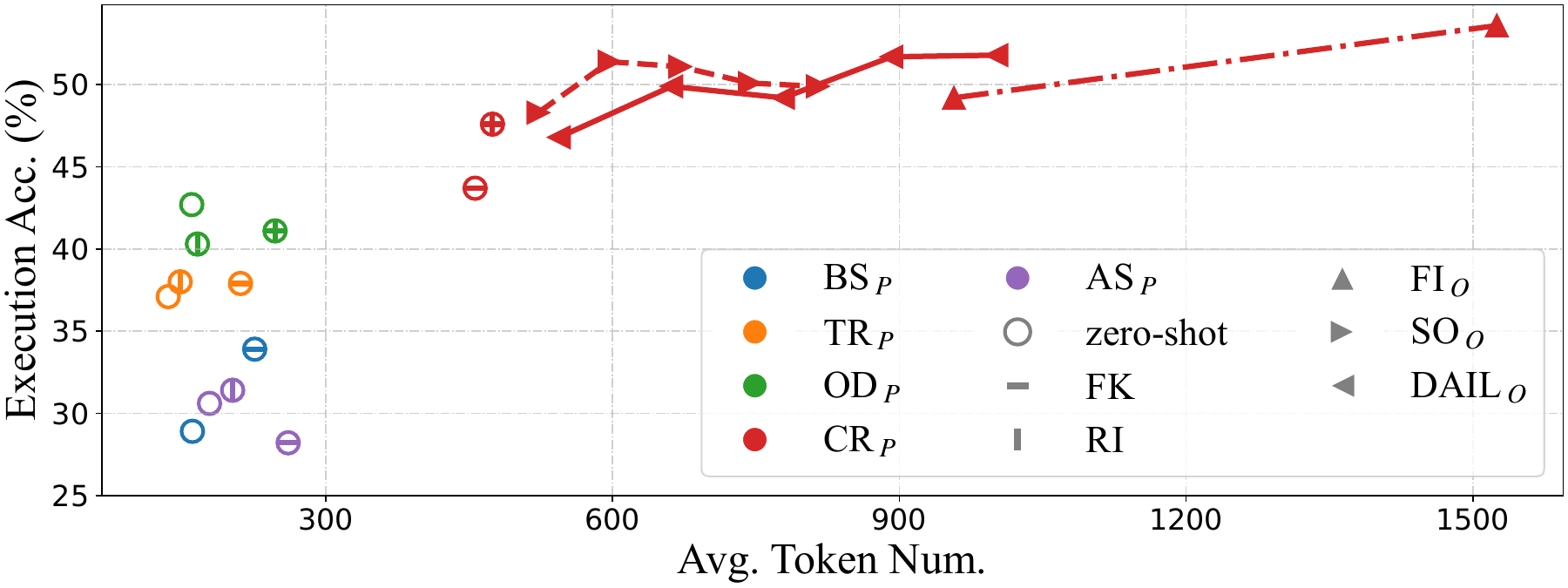}\label{fig:token:vicuna}}
    \subfigure[Averaged token number of each prompt and their time-consuming on Spider-dev.]{\includegraphics[width=0.49\linewidth]{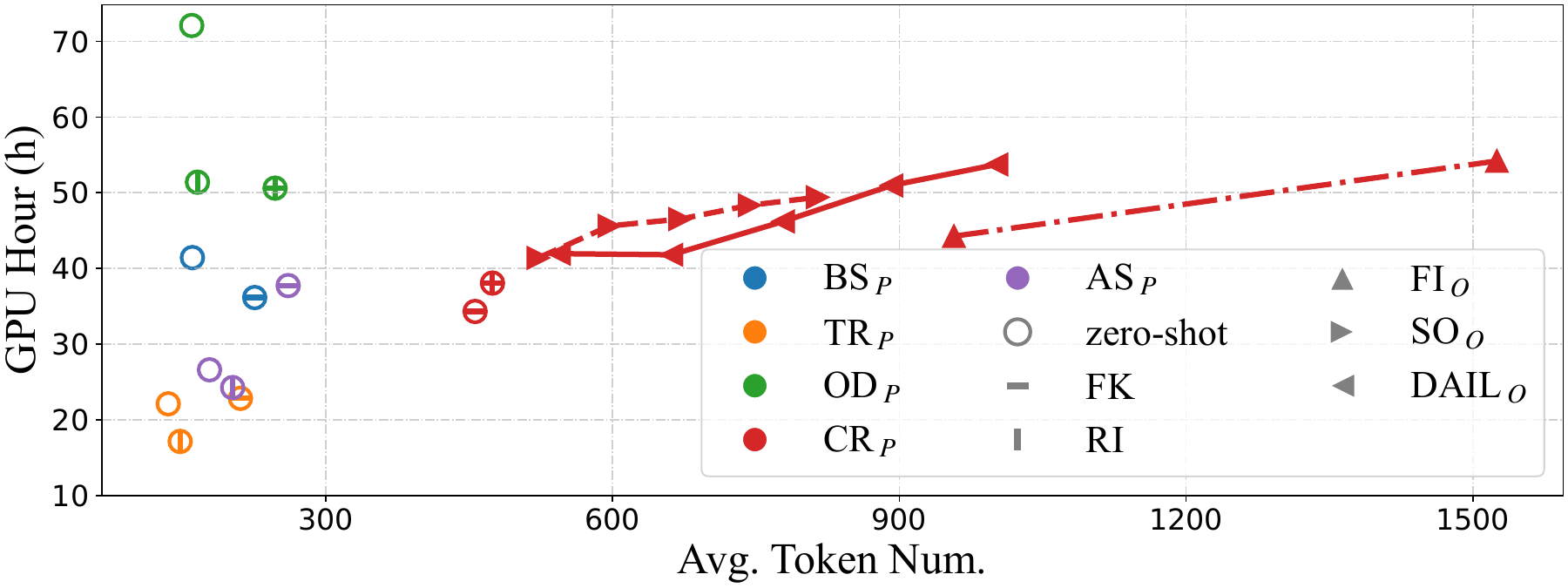}\label{fig:gpu_hour:vicuna}}
    
    \caption{The cost of our experiments with Vicuna-33B.}
    \label{fig:vicuna_cost}
\end{figure}

We estimate the cost of our experiments on the Vicuna-33B model, as shown in Figure~\ref{fig:vicuna_cost}. Unlike the cost of API calling, the number of output tokens also plays a significant role in the time consumption of Vicuna-33B during local inference. For instance, using the \openaiprompt (\abopenaiprompt) tends to elicit more time-consuming explanations accompanied by SQL queries. Note that the implication rule of ``with no explanation" not only improves performance but also saves time by suppressing Vicuna-33B from generating unnecessary explanations, as demonstrated in Figure~\ref{fig:vicuna_cost}. The average cost of GPU hours is 54.4 hours in each experiment.

\section{Leaderboard}
\subsection{Spider Leaderboard}
\begin{figure}[htb]
    \centering
    \includegraphics[width=0.8\linewidth]{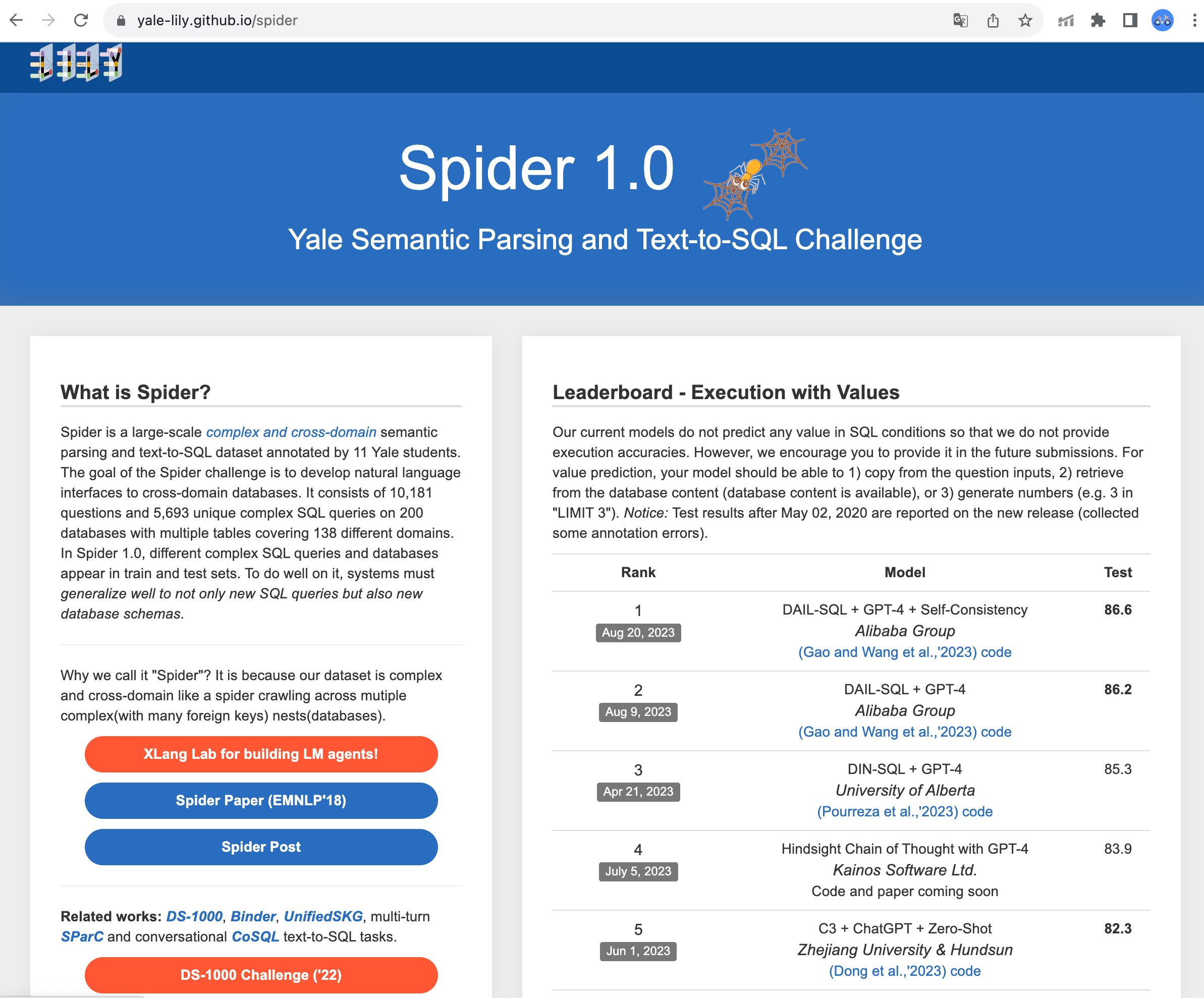}
    \caption{Current performance rank in Spider leaderboard. (Last accessed on 2023-09-19.)}
    \label{fig:leaderboard}
\end{figure}

\figref{fig:leaderboard} shows the performance rank in Spider~\cite{spider} leaderboard on Sep 19, 2023. 
In the leaderboard, our solution \ours with GPT-4 is reported to achieve $86.2\%$ execution accuracy; further, with self-consistency, our solution achieves $86.6\%$ execution accuracy, ranked as the first place.

\subsection{\revision{BIRD Leaderboard}}
\label{bird:leaderboard}

\begin{figure}[htb]
    \centering
    \includegraphics[width=0.75\textwidth]{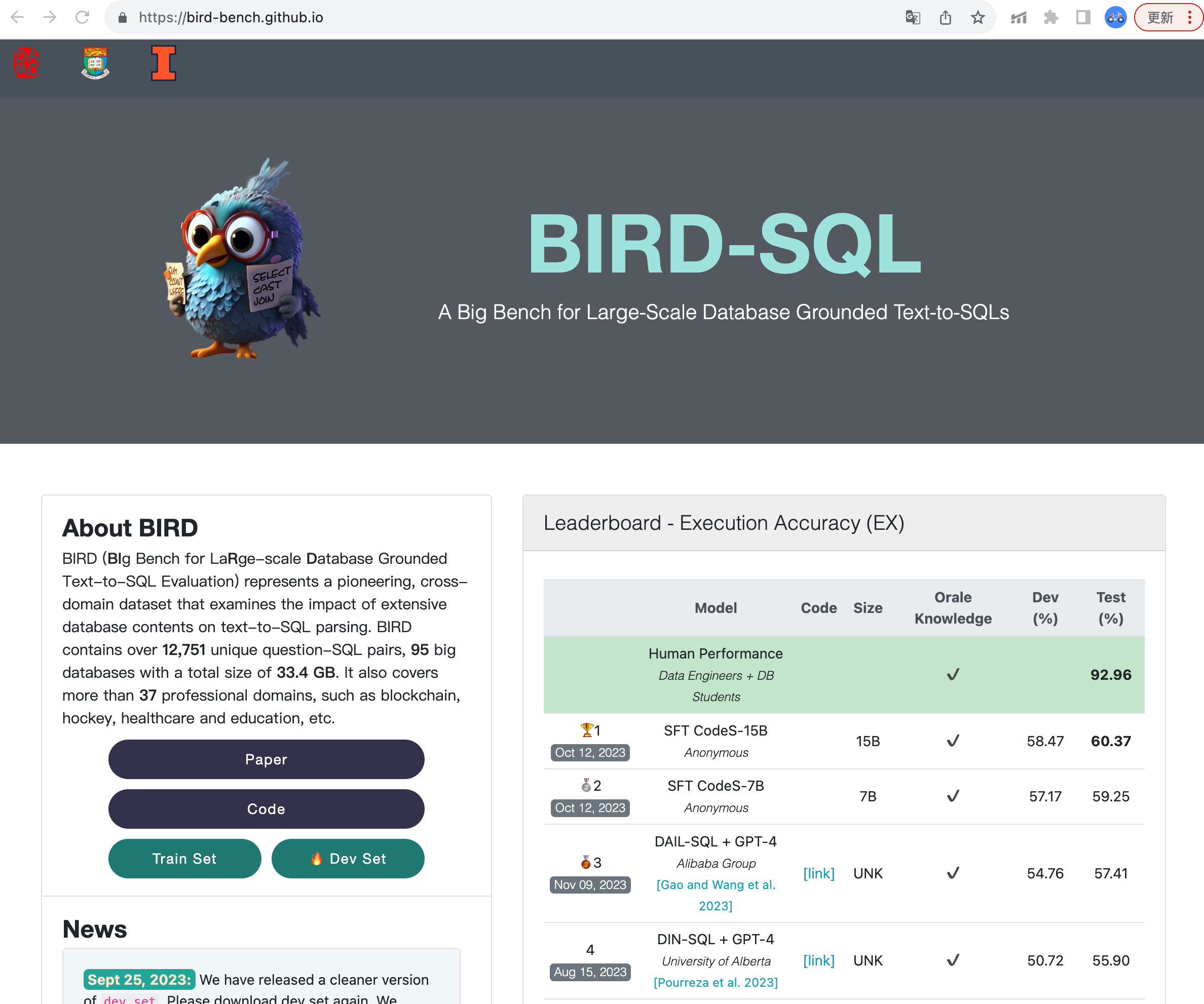}
    \caption{The leaderboard of BIRD at excution accuracy (EX).}
    \label{fig:bird_leaderboard}
\end{figure}

\figref{fig:bird_leaderboard} shows the performance rank in BIRD~\cite{DBLP:journals/corr/abs-2305-03111} leaderboard on Nov 09, 2023. We can observe that on the BIRD Benchmark, DAIL-SQL also outperforms the previous state-of-the-art method, DIN-SQL, by a remarkable $4.04\%$ in BIRD dev set and $1.51\%$ in BIRD test set in terms of execution accuracy. 

However, the additional challenge of BIRD lies in effectively leveraging the specific domain knowledge offered by the data set. As shown in Figure~\ref{fig:bird_leaderboard}, the top-1 and top-2 solutions have a SFT procedure, while DAIL-SQL have no such procedure and requires further design to incorporate extra knowledge for a more fair comparison. 

\end{document}